\documentclass[a4paper,11pt]{article}

\usepackage{jheppub} 

\usepackage{graphicx}
\usepackage{epstopdf}
\usepackage{amsmath}
\usepackage{enumitem}

\usepackage{epsfig}

\usepackage{subfigure}
\usepackage{longtable}
\usepackage{times}

\title{Evolution of Shock Structures and QPOs After Halting BHL Accretion onto Kerr Black Hole}

\author[a,1]{O. Donmez, \note{Corresponding author.}}

\affiliation[a]{College of Engineering and Technology, American
  University of the Middle East, Egaila 54200, Kuwait}

\emailAdd{orhan.donmez@aum.edu.kw}

  \abstract{   
  In this paper, we investigate for the first time the quasi-periodic oscillations (QPOs) that arise as a result of halting the Bondi-Hoyle-Lyttleton (BHL) accretion mechanism around a Kerr black hole. Unlike previous studies, which focused on shock cones or perturbed tori, we show that stopping BHL accretion leads to the formation of a new plasma structure. On this plasma structure, spiral shock waves are observed to develop. These shock waves are found to excite both low-frequency (LFQPOs) and high-frequency QPOs (HFQPOs). Two key results emerge from our numerical simulations. First, LFQPOs arise only when the halted flow is supersonic, identifying for the first time the velocity dependence of their formation. Second, nonlinear couplings that produce $3:2$, $2:1$, and $1:2:3$ resonances appear exclusively for rapidly rotating black holes ($a/M = 0.9$), directly linking spin to resonance states. In addition, it is demonstrated that HFQPOs are strongly influenced by the black hole spin. The generated QPOs are observed to be excited through the modes induced by the black hole spacetime curvature. The frequencies obtained from the numerical simulations align with observations of $GRS1915+105$, $XTEJ1550-564$, and $REJ1034+396$, thereby bridging stellar-mass and supermassive black hole systems within a unified framework. Thus, the newly uncovered physical mechanism associated with halted BHL accretion is capable of explaining both the persistence and the intermittency of QPOs across diverse astrophysical environments.
}


\begin{document} 
\maketitle
\flushbottom


\section{Introduction}
\label{Introduction}

BHL accretion is a process that describes how compact objects gravitationally capture matter in a non-spherical manner from their surrounding environment \cite{Bondi1952MNRAS, Bondi1, Edgar1}. The captured matter, initially dispersed into the interstellar medium by supernova explosions, is pulled inward by the strong gravitational force of a compact object such as a black hole, leading to the formation of an accretion disk around it. Consequently, this accretion mechanism is considered a viable model for the black holes interacting with the interstellar medium, $X-$ray binary systems, and active galactic nuclei (AGNs) \cite{ShapiroBook1985, Petrich1989, 2002apa..book.....F}. Theoretical studies have shown that the formation and structural evolution of an accretion disk in BHL accretion, along with possible oscillatory modes, depend on the black hole spin parameter, the physical properties of the resulting shock waves, and the angular momentum of spacetime in a strong gravitational field \cite{FontMNRAS1999, Font2000LRR, Foglizzo1999A&A, Foglizzo1}. Numerical simulations have shown that BHL accretion leads to the formation of shock cones around black holes, and the oscillatory behavior of these cones results in the generation of QPOs \cite{Donmez5, Koyuncu1, Ohsugi1, Wenrui1, CruzOsorio2020ApJ}. Observational evidence from $X-$ray binary systems and AGNs suggests that QPO formation around the black holes originates from the oscillations of the relativistic accretion disk and associated shock waves \cite{Hopkins2006, Remillard2006ARA&A, 2019NewAR..8501524I, Gatti2016}. General relativistic hydrodynamics (GRHD) and general relativistic magnetohydrodynamics (GRMHD) simulations of BHL accretion have not only provided insight into the shock wave structures but also revealed the fundamental QPO modes trapped within these shock regions \cite{Petrich1989, Donmez6, Narayan1, Donmez2024arXiv240701478D}. Recent numerical studies incorporating modified gravity theories have begun to explore deviations from general relativity \cite{AliOvgun1, AliOvgun2, AliOvgun3, AliOvgun4, Jav5}, investigating the extent to which QPO frequencies are affected by such modifications and their potential to explain observational data \cite{Donmezetal2022, CruzOsorio2023JCAP, Donmez2024Submitted, Donmez2024arXiv240216707D, Donmez2024Univ, AliQPO1}. Understanding the interplay between relativistic BHL accretion, shock oscillations, and QPO generation is crucial to interpreting high-energy astrophysical phenomena and probing the fundamental nature of gravity and the nature of dark matter in extreme environments \cite{Jav1, Jav2, Jav3, Jav4}.

The reduction or complete cessation of BHL accretion due to various physical reasons can significantly impact the structure of the shock cone that forms in a strong gravitational field. Consequently, the physical processes that arise from these changes can greatly influence QPOs and their frequencies. BHL accretion can be altered due to several physical mechanisms. For instance, as gas falls toward the black hole, it can accrete to a certain extent and create a potential barrier, which either significantly slows down or completely stops the flow of matter into the black hole \cite{2021PASJ...73.1429I, 2016MNRAS.463...63C}. Moreover, the infalling matter through the BHL process can encounter perturbations or obstacles, leading to turbulence or shock waves, which in turn can cause the accretion process to become highly chaotic or substantially reduced \cite{2015A&A...583A..90B, 2017MNRAS.471.3127C}. In strong gravitational fields, relativistic effects can alter the flow dynamics of the accumulated matter, thereby reducing the accretion rate. For example, the formation of a shock cone and the black hole spin parameter play a crucial role in decreasing the mass inflow rate in such strong gravitational environments \cite{2015A&A...583A..90B, CruzOsorio2020ApJ}. The interaction of the black hole with the plasma structure formed after the cessation of BHL accretion generates QPOs in the strong gravitational field, depending on the physical properties of both the black hole and the surrounding matter. Identifying the characteristics and understanding the nature of these oscillation modes, which are known as LFQPOs and HFQPOs, will make a significant contribution to the literature.

LFQPOs, calculated from variations in $X-$ray data, are generally observed in black hole binary systems. That is, they arise as a result of the interaction between the black hole and matter, as well as the structure of the accretion disk formed by the infalling material toward the black hole. These QPOs occur both around stellar-mass black holes and around supermassive black holes at the centers of AGNs. For stellar-mass black holes, LFQPO frequencies typically fall within the $0.1–30$ Hz range. Observing these frequencies not only allows us to understand the dynamic structure of the disk near the event horizon but also helps reveal the interaction between matter and black holes in strong gravitational fields \cite{2005ApJ...629..403C, Ingram2009MNRAS, 2019NewAR..8501524I}. LFQPOs are classified into Type-A, Type-B, and Type-C, based on their frequency, coherence, and spectral properties \cite{1999ApJ...514..939W, Motta2014MNRAS, 2016AN....337..398M}. Among these, Type-C QPOs are typically associated with Lense-Thirring precession in the accretion disk and are commonly observed in the hard and intermediate states of $X-$ray binary systems. Lense-Thirring precession is entirely related to the curvature of spacetime induced by the black hole spin \cite{Ingram2009MNRAS}. Due to this characteristic, LFQPOs provide important insight into general relativistic effects \cite{1998ApJ...492L..59S}. LFQPOs have been observed in well-known black hole sources such as $GRS 1915+105$, $GRO J1655-40$, $XTE J1550-564$, and $H1743-322$ \cite{Remillard2006ARA&A, BelloniMNRAS2012}. Furthermore, QPO-like signals have been detected in the AGN $RE J1034+396$, where a QPO frequency in the mHz range has been identified \cite{2011MNRAS.417..250M}. Studying LFQPOs is particularly important because they provide information about strong gravitational fields and the regions near black holes. However, their exact physical origin remains uncertain \cite{1999A&A...349.1003T, 2006ApJ...637L.113S}. Therefore, numerical studies of LFQPOs, investigating their physical causes and comparing the results with observations, are essential to advance the scientific literature.

The precessional motions and sudden changes in plasma structures around the stellar-mass and the supermassive black holes, caused by different physical processes, lead to the formation of HFQPOs. These QPOs are observed in both binary systems and Active Galactic Nuclei (AGNs). The oscillation frequencies of these QPOs typically range between $40–450$ Hz for stellar-mass black holes, whereas for supermassive black holes, they vary from mHz to $\mu$Hz, depending on the black hole mass. It is well established that these QPOs originate from regions close to the event horizon, where strong gravitational fields dominate \cite{Remillard2006ARA&A}. The detection or observation of QPOs not only helps in understanding the physical properties of black holes but also provides insights into the behavior of matter in strong gravitational fields and the dynamic structure of plasma. Observational studies show that HFQPOs often appear in resonance ratios such as $3:2$. Examples include $GRO J1655-40$ ($300$, $450$ Hz), $XTE J1550-564$ ($184$, $276$ Hz), and $GRS 1915+105$ ($41$, $67$ Hz) \cite{Remillard1999ApJ, Strohmayer2001ApJ, Misra2004,2012MNRAS.427..595M, Motta2014MNRAS, Varniere_2018, Liu_2021, Majumder_2022, Motta2023}. Theoretical explanations suggest that these resonance states arise from different oscillation modes, epicyclic motion, or relativistic precession effects \cite{Abramowicz2001A&A, Ingram2019}. Despite their astrophysical significance, HFQPOs have been observed in only a few sources, and even in those cases, they appear intermittently before disappearing. To confirm the accuracy of these frequencies observationally, long-term X-ray observations are required. By analyzing HFQPO data, we can extract crucial information about black hole mass, spin parameters, and accretion mechanisms \cite{Zhang_2017}. To improve the precision of these findings, long-duration observations are essential. Therefore, studying HFQPOs should not rely solely on observational efforts; instead, numerical and theoretical studies are also crucial to uncovering their physical nature. Through these combined efforts, we can gain a deeper understanding of the black hole environments and test the strong gravitational fields more effectively.

In this study, we investigate the physical structure of the plasma that forms in the strong gravitational field around the black hole after the shock cone has developed due to BHL accretion and when BHL accretion is later halted due to specific physical phenomena. We examine how the nature of the newly formed plasma changes depending on the black hole spin parameter and the asymptotic velocity used in the BHL accretion process. Additionally, we analyze the precession motion of the plasma in the strong gravitational field and the resulting QPO frequencies caused by spacetime curvature. Finally, we compare these QPOs with observational results, theoretical expectations, and previous QPO calculations in the literature, discussing the contribution of our findings to the field.

The structure of the paper is organized as follows: In Section \ref{Mat_frame}, the GRH equations and their numerical solution methods are presented, together with the Kerr spacetime metric. In addition, the physical variables used as initial conditions in the numerical simulations, as well as the boundary conditions, are summarized. In Section \ref{AstroMot}, various astrophysical scenarios related to the slowing down or complete cessation of the BHL mechanism are introduced, which explains the necessity of numerical simulations. In Section \ref{NumRes}, the numerical solutions of the GRH equations with the appropriate initial conditions are discussed. The behavior of the system is analyzed based on different black hole spin parameters and the asymptotic velocity of the infalling gas. The numerical results are presented for both the rest-mass density distributions in the equatorial plane and the mass accretion rate, followed by power spectrum density analyses. The dependence of HFQPOs and LFQPOs on various parameters is discussed in Section \ref{low_high_QPOs}. In Section \ref{comparision}, the numerical results obtained in this study are compared with the observed QPOs, highlighting the strong agreement between the numerical and observational findings for certain astrophysical sources. In Section \ref{compare_previous}, a comparison is made between perturbed tori and shock cone oscillations previously studied and the QPOs found in this work. The discussion emphasizes that the results of this study provide a better explanation for certain microquasar observations compared to existing models in the literature. Finally, Section \ref{Conclusions} is devoted to the conclusions of our work. Throughout the paper, the Latin indices $i$ and $j$ range from 1 to 3, while the indices $a$, $b$, and $c$ take values from 0 to 3. That is, $i$ and $j$ represent spatial indices, while $a$, $b$, and $c$ correspond to spacetime indices. In this study, we adopt geometrized units with $G=c=1$ to express the results in terms of the mass of the black hole. This choice facilitates the application of our numerical results to both stellar-mass and supermassive black holes.


\section{Mathematical Framework}
\label{Mat_frame}

\subsection{GRHD Equations}
\label{GRHD_Eq}

In this paper, we numerically solve the GRH equations using the Kerr metric while considering the ideal gas assumption to model the formation of the shock cone around the black hole due to BHL accretion and the plasma structure that emerges after the cessation of the BHL mechanism in a strong gravitational field. To solve the GRH equations using high-resolution numerical methods, we first express them in conservation form. The GRH equations in conservation form are \cite{1997ApJ...476..221B},

\begin{eqnarray}
  \frac{\partial U}{\partial t} + \frac{\partial F^r}{\partial r} + \frac{\partial F^{\phi}}{\partial \phi}
  = S.
\label{GREq1}
\end{eqnarray}

\noindent
Here, $U$, $F^r$, $F^{\phi}$ and $S$ represent the conserved variable vector, the flux vector in the radial direction, the flux vector in the azimuthal direction and the source vector, respectively. These vector variables are expressed in terms of the primitive parameters ($\rho$, $v^i$, $p$) as follows:

\begin{eqnarray}
  U =
  \begin{pmatrix}
    D \\
    S_j \\
    \tau
  \end{pmatrix}
  =
  \begin{pmatrix}
    \sqrt{\gamma}W\rho \\
    \sqrt{\gamma}h\rho W^2 v_j\\
    \sqrt{\gamma}(h\rho W^2 - P - W \rho)
    \end{pmatrix},
\label{GREq2}
\end{eqnarray}

\noindent
where $v_j$ represents the three-velocity, which is expressed as $v^i = u^i/W + \beta^i/\alpha$ in terms of the four-velocity and other variables. $\alpha$ and $\beta^i$ are the lapse function and the shift vector, respectively. They are defined below for the Kerr metric. The three-velocity is measured in the Eulerian reference frame. The indices appearing here can be written in upper or lower form using the three-metric $\gamma_{i,j}$. $W = (1 - \gamma_{a,b}v^i v^j)^{1/2}$ denotes the Lorentz factor of the fluid, while $\gamma$ represents the determinant of the spatial metric. $h = 1 + \epsilon + P/\rho$ represents the specific enthalpy. Based on the defined primitive and conservative variables, the fluxes and source terms are given as follows \cite{Font2000LRR}:

\begin{eqnarray}
  {F}^r =
  \begin{pmatrix}
    \alpha\left(v^r - \frac{\beta^r}{\alpha}\right)D \\
    \alpha\left(\left(v^r - \frac{\beta^r}{\alpha}\right)S_j + \sqrt{\gamma}P\delta^r_j\right)\\
    \alpha\left(\left(v^r - \frac{\beta^r}{\alpha}\right)\tau  + \sqrt{\gamma}P v^r\right)
    \end{pmatrix},
\label{GREq3}
\end{eqnarray}

\begin{eqnarray}
  {F}^{\phi} =
  \begin{pmatrix}
    \alpha\left(v^{\phi} - \frac{\beta^{\phi}}{\alpha}\right)D \\
    \alpha\left(\left(v^{\phi} - \frac{\beta^{\phi}}{\alpha}\right)S_j + \sqrt{\gamma}P\delta^{\phi}_j\right)\\
    \alpha\left(\left(v^{\phi} - \frac{\beta^{\phi}}{\alpha}\right)\tau  + \sqrt{\gamma}P v^{\phi}\right)
    \end{pmatrix},
\label{GREq4}
\end{eqnarray}

\noindent and,

\begin{eqnarray}
  \vec{S} =
  \begin{pmatrix}
    0 \\
    \alpha\sqrt{\gamma}T^{ab}g_{bc}\Gamma^c_{aj} \\
    \alpha\sqrt{\gamma}\left(T^{a0}\partial_{a}\alpha - \alpha T^{ab}\Gamma^0_{ab}\right)
   \end{pmatrix}, 
\label{GREq5}
\end{eqnarray}

\noindent where $\Gamma^c_{ab}$ is the Christoffel symbol.

During defining the behavior of matter around the black hole, we assumed that it behaves as a perfect fluid. Consequently, the equation of state is given as $P = (\Gamma - 1)\rho\epsilon$. Here, $\epsilon$ and $\Gamma$ represent the specific internal energy and the adiabatic index of the mater, respectively. In addition to this, solving the GRH equations numerically requires defining the four-metric $g_{ab}$ that describes spacetime, along with the associated three-metric $\gamma_{i,j}$ and Christoffel symbols $\Gamma^c_{ab}$, to determine the structure of the disk and other astrophysical phenomena around the black hole. In this study, we use the Kerr black hole metric in Boyer-Lindquist coordinates. This metric is given as follows \citep{Donmez6}:

\begin{eqnarray}
  ds^2 = -\left(1-\frac{2Mr}{\sum^2}\right)dt^2 - \frac{4Mra}{\sum^2}sin^2\theta dt d\phi
  + \frac{\sum^2}{\Delta_1}dr^2 + \sum^2 d\theta^2 + \frac{A}{\sum^2}sin^2\theta d\phi^2,
\label{GREq6}
\end{eqnarray}

\noindent where $\Delta_1 = r^2 - 2Mr +a^2$, and
$A = (r^2 + a^2)^2 - a^2\Delta sin^2\theta$.
The lapse function and the shift vector of the Kerr metric are
$\alpha = (\sum^2 \Delta_1/A)^{1/2} $ and $\beta^i = (0,0,-2Mar/A)$.


\subsection{Initial Conditions}
\label{Init_Cond}

Since we are modeling the interaction between the black hole and the surrounding plasma in the two-dimensional equatorial plane, we set $\theta=\frac{\pi}{2}$. One of the main reasons for choosing 2D simulations is computational efficiency. This allows us to run each model on very long timescales. In this paper, in order to analyze the quasi-periodic oscillations (QPOs) that arise after the BHL mechanism is halted, the code is evolved up to $80000M$. At the same time, by using 2D simulations, we were able to employ much higher resolution, thereby minimizing the impact of numerical errors and spurious oscillations. As demonstrated in \cite{2011MNRAS.417.2899Z}, most of the key features of the shock cone produced by the BHL mechanism can already be captured in 2D simulations. On the other hand, as shown in the literature, 3D simulations generally cannot achieve such high resolution due to computational cost and runtime limitations \citep{2022ApJ...933L...9G}. Of course, since we did not perform 3D simulations here, vertical instabilities, MRI-driven turbulence, and jet collimation cannot be addressed in this study. However, once the 3D GRMHD code we are currently developing is operational, we will have the opportunity to model these effects and compare the physical outcomes.

In this study, we first simulate the formation of a shock cone in a strong gravitational field due to the interaction between the black hole and infalling matter through BHL accretion. The shock cone reaches a quasi-steady state at approximately $t=2000$M. It is observed that this steady-state condition persists until around $t=6000$M. After this period, BHL accretion is halted, and the subsequent interaction between the black hole and the surrounding matter leads to the formation of a new plasma structure. For rapidly rotating black holes ($a/M=0.9$), the newly formed plasma is believed to reach a stable configuration between $t=40000$M and $t=80000$M. During this interval, we compute the mass accretion rate and perform a power spectrum analysis based on this accretion behavior. In the case of slowly rotating ($a/M=0.5$) and non-rotating ($a/M=0$) black holes, the plasma is assumed to attain stability between $t=26000$M and $t=46000$M, and similar calculations are conducted.

Initially, during the modeling of the shock cone formation through BHL accretion, the infall of matter towards the black hole is governed by the asymptotic velocity $V_{\infty}$. The different values of $V_{\infty}$ used in the numerical simulations can be seen in Table \ref{Inital_Con_1}. The matter is injected from the outer boundary towards the black hole with the following radial and azimuthal velocities \cite{Font2000LRR}:

\begin{eqnarray}
V^r= \sqrt{\gamma^{rr}}V_{\infty}cos(\phi), \nonumber \\
V^{\phi}= -\sqrt{\gamma^{\phi\phi}}V_{\infty}sin(\phi).
\label{GREq7}
\end{eqnarray}

Throughout the evolution during the creation of the shock cone, matter is allowed to fall toward the black hole with the velocities given in Eq.\ref{GREq7}. The rest-mass density and pressure of the infalling matter are chosen such that the sound speed is $C_s=0.1$. To achieve this, we set $\rho=1$. Consequently, for $C_s=0.1$, the pressure is determined using the ideal gas equation of state as $P=C_s^2\rho(\Gamma-1)/[\Gamma(\Gamma-1)-C_s^2\Gamma]$. In all numerical calculations, the adiabatic index is taken as $\Gamma=4/3$.

We neglect the effects of magnetic fields and perform purely hydrodynamic simulations. The primary reason for this simplification is that, since we focus on the region very close to the black hole ($r < 10M$), the gravitational timescale is much shorter than the magnetic timescale, allowing us to reasonably ignore the influence of magnetic fields. Previous studies have shown that magnetic fields affect the instabilities of shock dynamics, the structure of jets formed in strong gravitational fields, and the overall accretion flow dynamics \citep{2012MNRAS.426.3241N}. Our goal here is to reveal the type of effects produced when BHL accretion is halted within a GRHD framework. Undoubtedly, including magnetic fields in the calculations would introduce additional effects on the mechanisms at play. For example, magnetic pressure and tension could stabilize or destabilize the shock cone, alter the morphology of the spiral shock waves that we observe in our simulations, and change both the detectability and the actual values of the resulting QPO frequencies. Moreover, the magneto-rotational instability (MRI) could induce turbulence in the accretion flow, which in turn would influence the QPOs. Therefore, we believe that our assumptions are reasonable since the numerical simulations presented here provide not only fundamental physical insight but also capture key processes in a strong gravitational field. However, once we complete the GRMHD code we are currently developing, we will be able to assess the impact of magnetic fields on these phenomena and compare the resulting physical outcomes.

The computational domain consists of $1024$ points in the radial direction and $512$ points in the azimuthal direction. The radial coordinate extends from $r_{min}=2.3M$ to $r_{max}=100M$, while the azimuthal coordinate spans from $0$ to $2\pi$. In both this study and my previous works, it has been demonstrated that the resulting QPO oscillations are independent of the resolution. In the radial direction, outflow boundary conditions are applied both near the black hole horizon and at the outer boundary. If gas is injected toward the black hole in the upstream region as part of the BHL accretion process, the physical parameters of the injected matter are automatically copied into the ghost zones. The outflow boundary conditions help prevent numerical artifacts from propagating into the region where the disk or plasma is present. In the azimuthal direction, the periodic boundary condition is implemented.

\begin{table}
\footnotesize
\caption{
The initial physical variables used are as follows: the black hole spin parameter $a/M$, the asymptotic velocity of the matter injected from the outer boundary $V_{\infty}/c$, the Mach number $M=\frac{V_{\infty}}{C_s}$, and the rest-mass density  of the matter injected from the outer boundary $\rho/M^2$, respectively. In all initial conditions, the speed of sound is taken as $C_s=0.1$. However, the pressure of the infalling matter are calculated based on the speed of sound and the rest-mass density.
}
 \label{Inital_Con_1}
\begin{center}
  \begin{tabular}{ccccccccc}
    \hline
    \hline

    $$& $a/M$   & $$ &$V_{\infty}/c$ & $$ & $M$ & $$ & $\rho M^2$ & $$   \\ 
    \hline
    $$ & $0.9$& $$ & $0.1$ & $$ & $1$ & $$ & $1$ & $$    \\
    $$ & $0.9$& $$ & $0.2$ & $$ & $2$ & $$  & $1$ & $$   \\
    $$ & $0.9$& $$ & $0.3$ & $$ & $3$ & $$ & $1$ & $$    \\
    $$ & $0.9$& $$ & $0.4$ & $$ & $4$ & $$ & $1$ & $$    \\
     \hline   
    $$ & $0.5$& $$ & $0.2$ & $$ & $2$ & $$ & $1$ & $$    \\
    $$ & $0.0$& $$ & $0.2$ & $$ & $2$ & $$  & $1$ & $$   \\
 
    \hline
    \hline
  \end{tabular}
\end{center}
\end{table}
%


\section{Astrophysical Motivation}
\label{AstroMot}

BHL accretion is the process by which matter expelled into empty space due to supernova explosions falls towards a black hole, forming a plasma around it. Numerical calculations, conducted with Kerr and Modified Gravity models, have shown that this accreted matter forms a shock cone around the black hole. These calculations also provide strong predictions and evidence that this cone is very likely to serve as a physical mechanism capable of explaining certain observational results related to QPOs \citep{Donmez5, Donmez4, Donmez3, Donmez_EGB_Rot, Donmezetal2022, Donmez2023arXiv231013847D, Donmez2024Submitted, Donmez2024arXiv240701478D, Donmez2024arXiv240216707D, Donmez2024Univ}. On the other hand, there are scenarios in which BHL accretion may come to an end. For instance, if the density of matter falling towards the black hole decreases due to stellar winds, this could significantly reduce, though not completely stop, BHL accretion \citep{2009MNRAS.398.2152D, Blondin_2012}. Secondly, a jet that may form around the black hole could cause matter to be ejected out of the system rather than falling towards the black hole, thus clearing the area around it of matter \cite{2007Ap&SS.307...17C, 2013EPJWC..6101005F}. Third, the presence of another compact object passing near the accreted matter could exert a strong gravitational influence, preventing further BHL accretion toward the black hole \citep{1999A&A...346..861R}. The clearing of the area around the black hole results in a powerful outward emission from the black hole. This type of unexpected situation was observed in \citep{2020MNRAS.497.1925G, cfa2022}. This strong outburst causes accretion to either completely stop or be delayed. 
The slowing down or halting of BHL accretion due to various physical reasons can lead to instability in the physical mechanism around the black hole and result in episodic accumulation phases. This irregular behavior around the black hole may appear in the observed X-ray data, potentially causing variations and irregularities in the observed QPO frequencies. Therefore, it can be used to explain the low- and high-frequency observational results seen in sources such as GRS 1915 + 105 \cite{Morgan_1997, Strohmayer2001ApJ, Misra2004, Belloni2013MNRAS, Rawat2019ApJ, Motta2023, Chauhan2024MNRAS}, XTE J1550-564 \cite{Remillard1999ApJ, Miller_APJ_2001, Homan2003ApJ, Motta2022MNRAS, Rink2022MNRAS}, GRO J1655-40 \citep{Remillard1999ApJ, Strohmayer_APJ_2001, 2012MNRAS.427..595M, Rink2022MNRAS, Remillard2002APS}, and H1743-322 \citep{Remillard2006ApJ, 2006ApJ...637.1002R, 2012ApJ...754L..23A}. Although this scenario may seem naturally challenging, modeling it theoretically could numerically reveal the QPO behavior observed in different black holes.


\section{Numerical Results}
\label{NumRes}

As seen in Section \ref{AstroMot}, BHL accretion can be either slowed down or completely halted due to various astrophysical reasons such as density and velocity gradients around the black hole, hydrodynamical, thermal instabilities, stellar winds, etc. This can significantly affect both the dynamic structure and even the existence of the shock cone mechanism formed around the black hole. Changes in the dynamic structure of the shock cone, or its complete disappearance, may lead to variations in the observed QPOs or the emergence of different QPO frequencies. In order to investigate these scenarios, we consider the variations in the black hole rotation parameter and asymptotic velocity, which strongly influence the dynamic structure of BHL accretion.

The QPO frequencies obtained in this study may originate from the fundamental modes already known in the literature \citep{CruzOsorio2023JCAP, Donmez2024arXiv240216707D}, or from their nonlinear couplings \citep{LL1976}. Nonlinear couplings occur when radial and azimuthal modes of the plasma, generated by spiral shock waves around the black hole, or modes associated with Lense–Thirring precession due to black hole rotation \citep{Ingram2009MNRAS}, interact to produce new frequencies through sums, differences, or integer multiples of the base frequencies. In the strong gravitational field of a rotating black hole, such couplings naturally arise as a consequence of differential rotation and the interaction between matter and the curved spacetime. For example, the radial epicyclic mode can couple with the azimuthal epicyclic mode, or with the Lense–Thirring precession mode, to generate resonances such as $2:1$ or $3:2$. These resonances are most often observed as HFQPOs. The presence of strong black hole spin enhances frame dragging and shear, which increases the likelihood of mode–mode interactions. As given in the following, this explains why in our results nonlinear couplings appear for rapidly rotating black holes ($a/M = 0.9$), but vanish for moderate or zero spin cases. Theoretically, such mode interactions are consistent with the nonlinear resonance interpretation of QPOs \citep{Abramowicz2001A&A, Rezzolla2003MNRAS}. Therefore, as discussed in detail below and confirmed by our numerical calculations, nonlinear coupling modes can be regarded as one of the primary mechanisms responsible for the generation of HFQPOs.

\subsection{Rapidly Rotating Black Hole with $a/M=0.9$}
\label{Rap_Num}

\begin{figure*}
  \vspace{1cm}
  \center
     \psfig{file=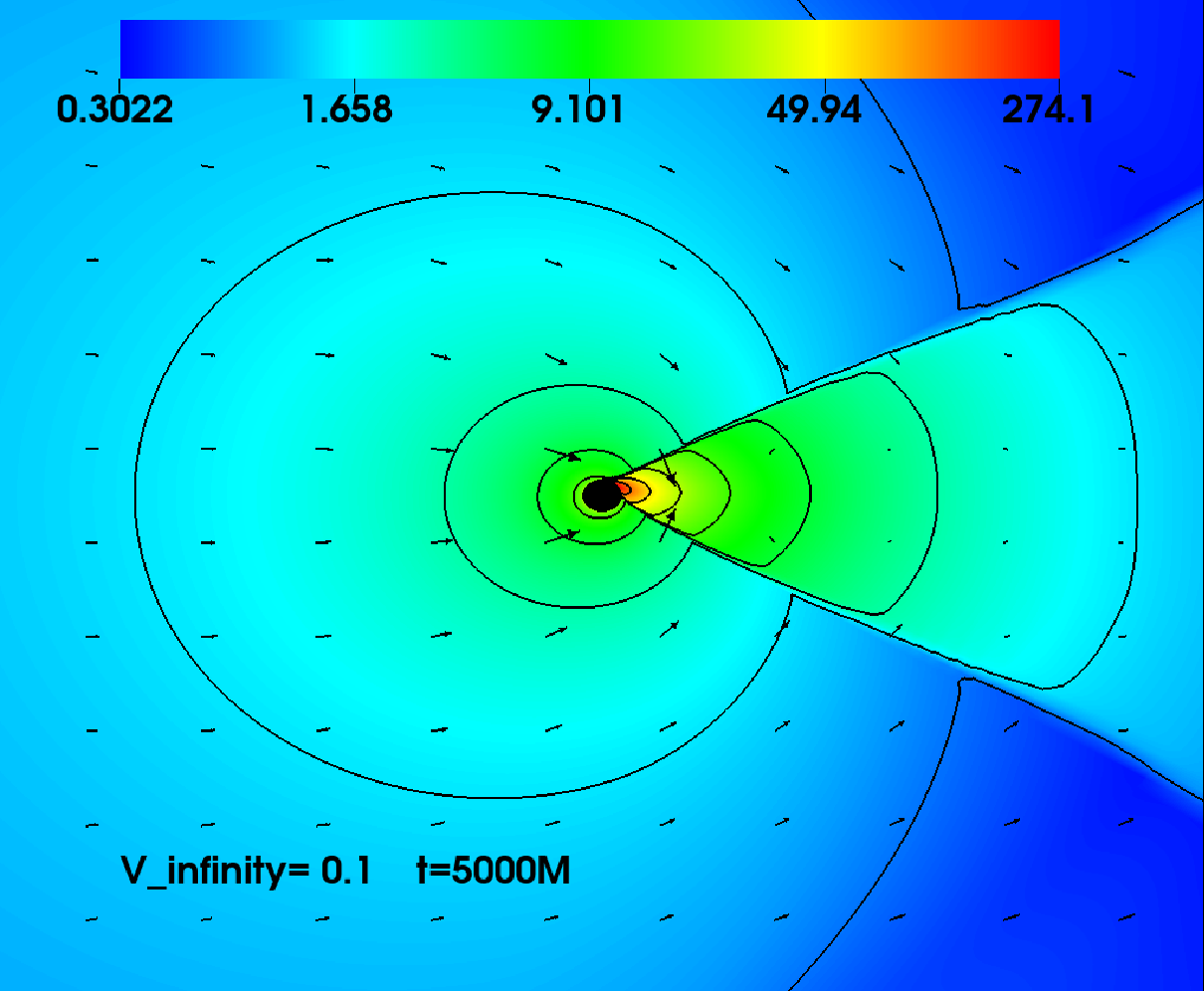,width=7.0cm, height=5.0cm}
     \psfig{file=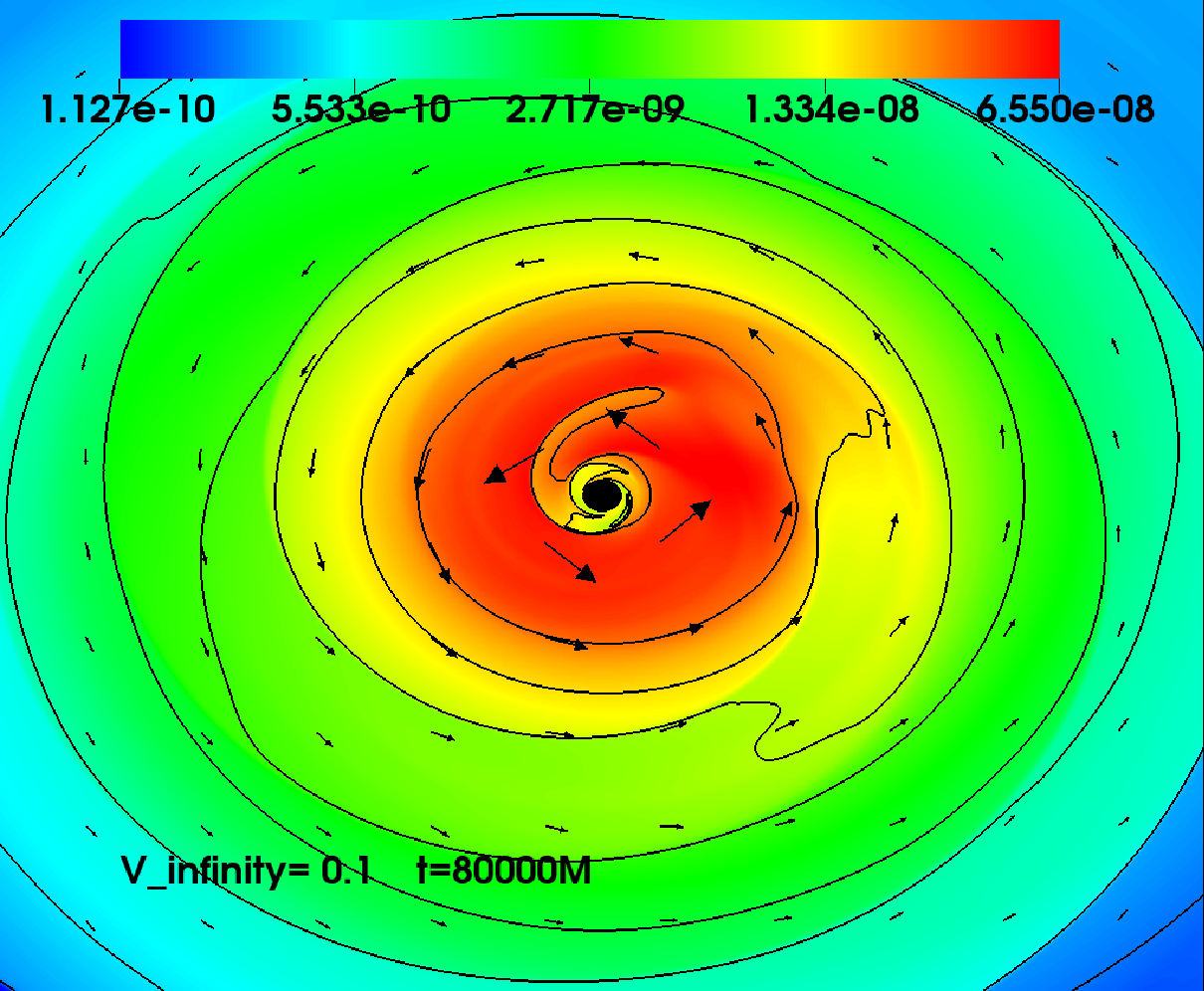,width=7.0cm, height=5.0cm}
     \psfig{file=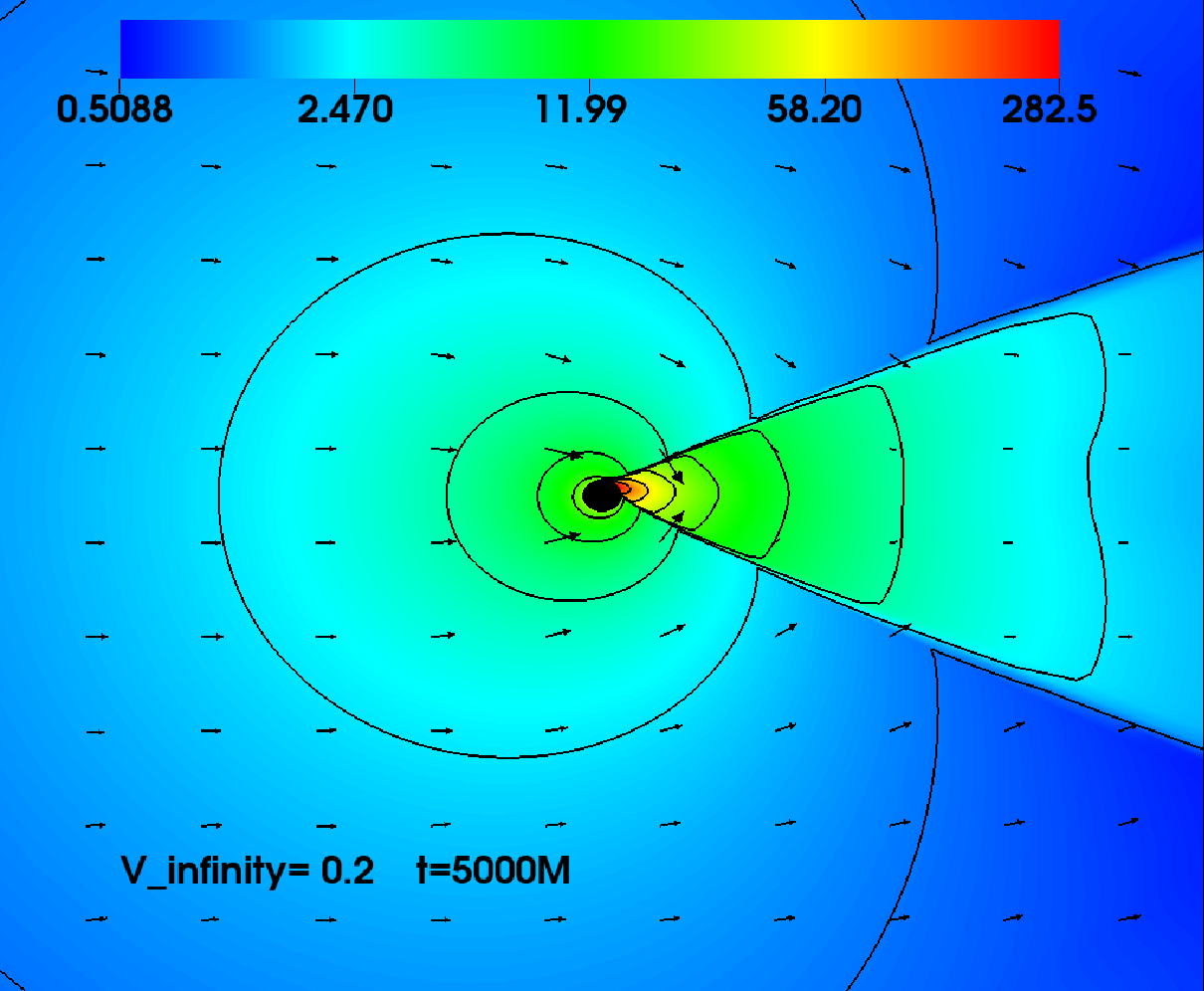,width=7.0cm, height=5.0cm}
     \psfig{file=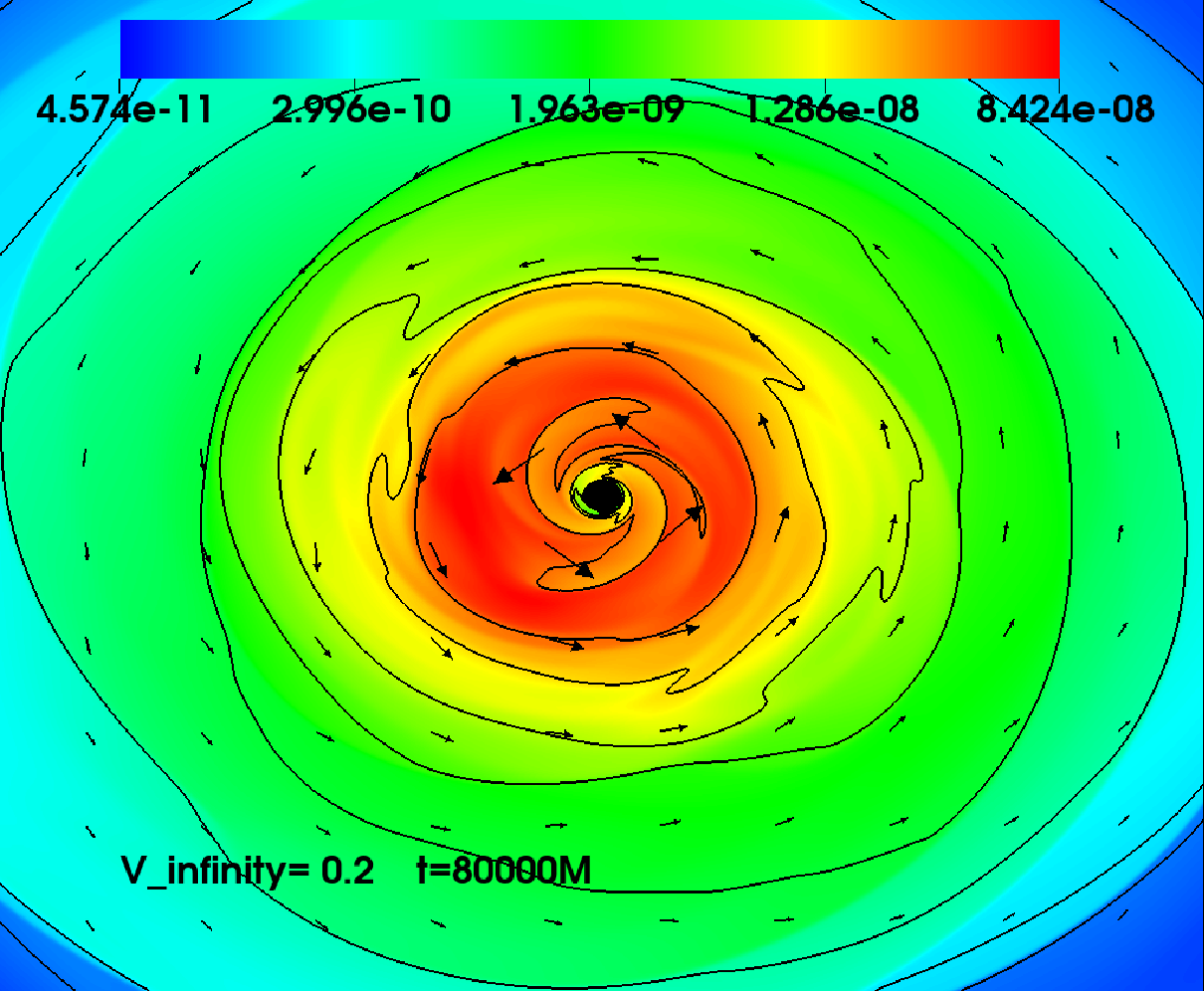,width=7.0cm, height=5.0cm}
     \psfig{file=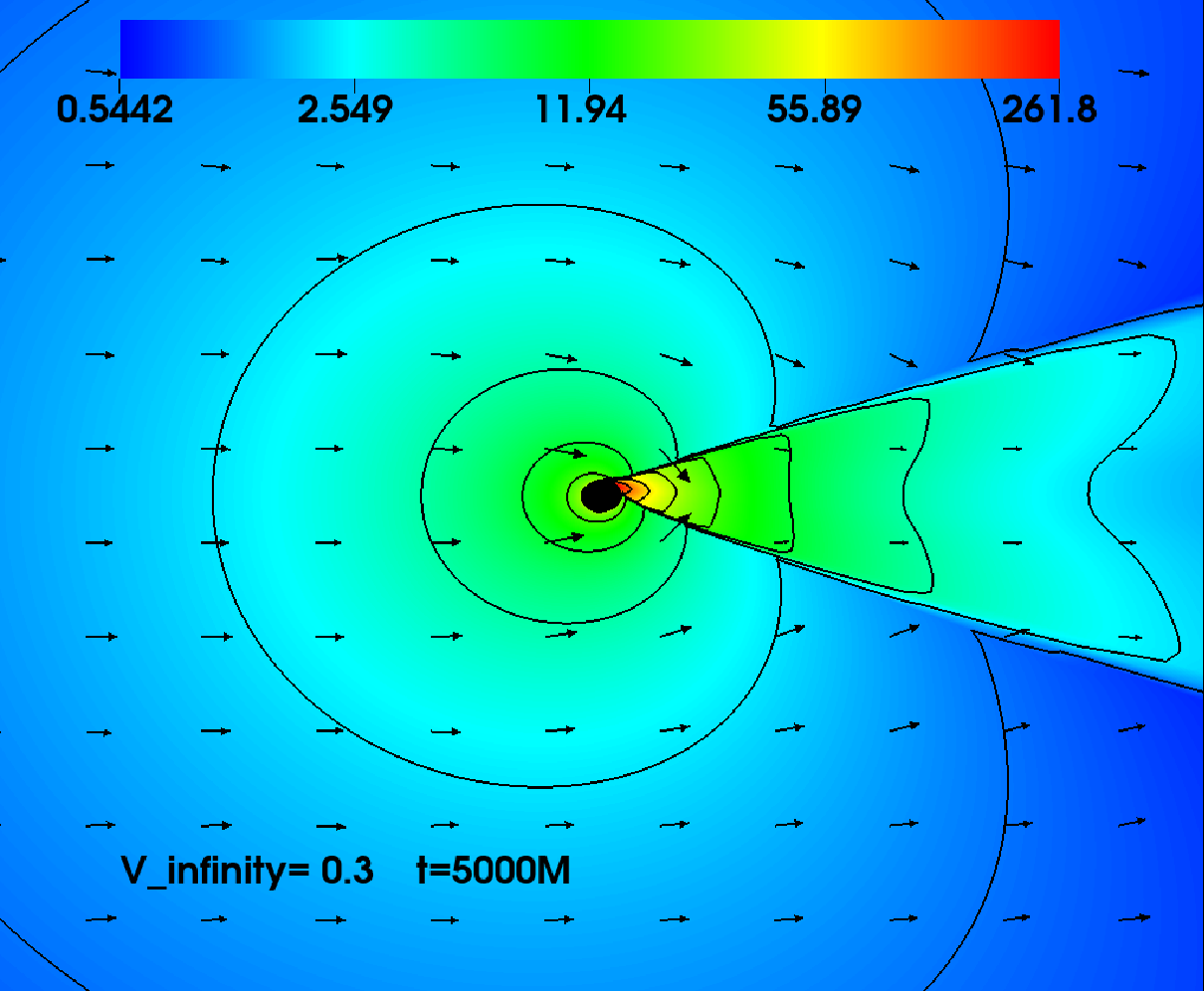,width=7.0cm, height=5.0cm}
     \psfig{file=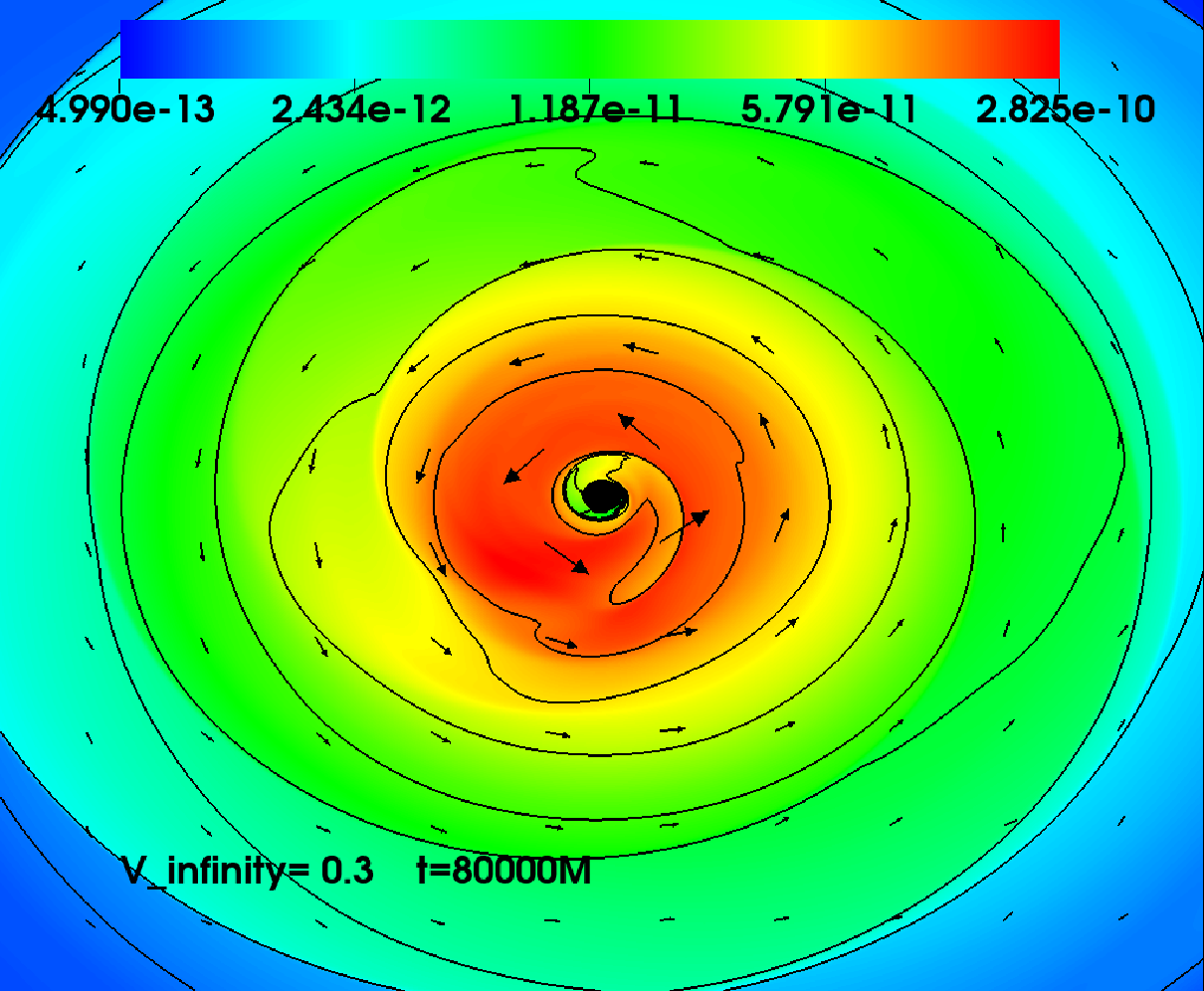,width=7.0cm, height=5.0cm}
     \psfig{file=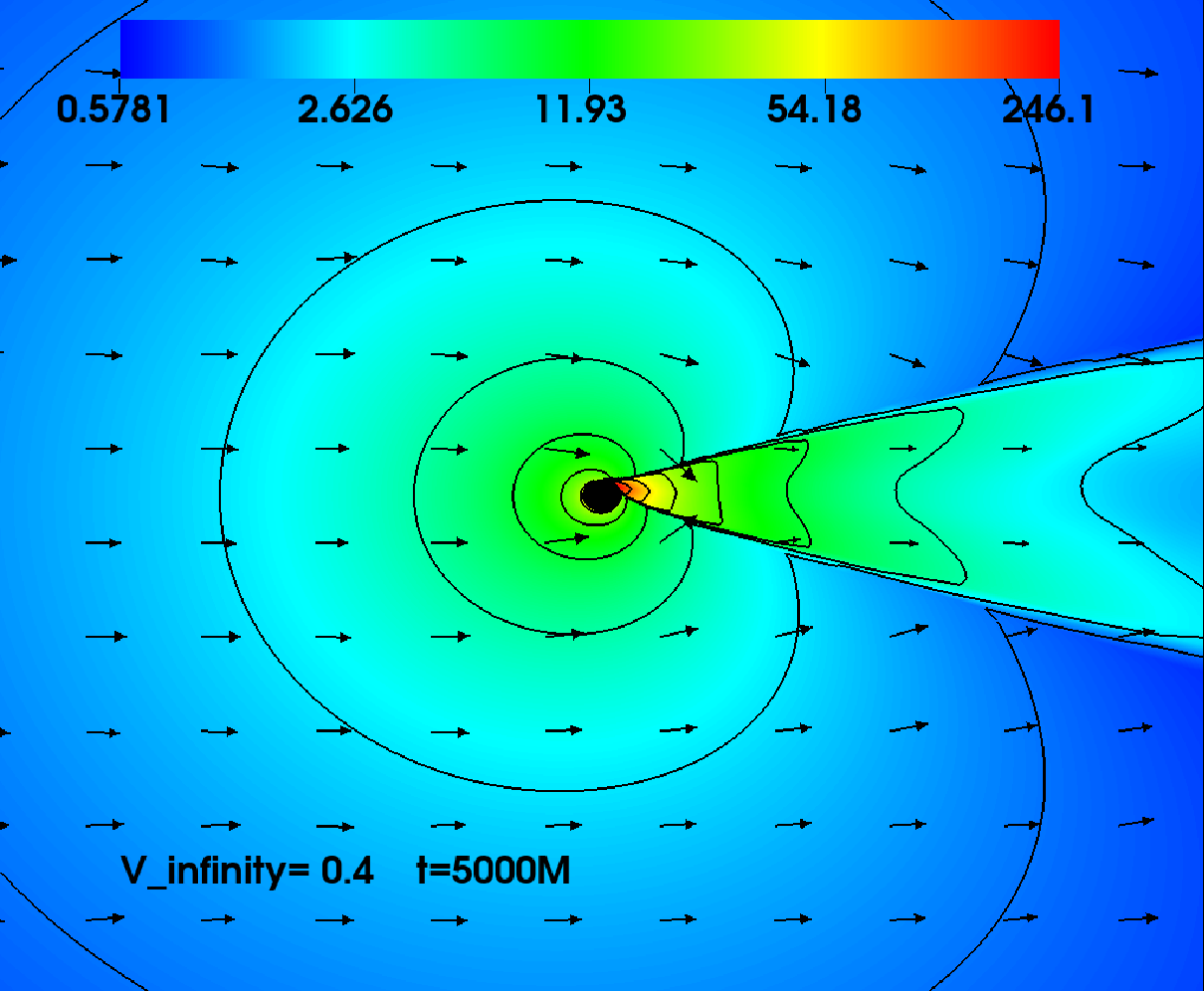,width=7.0cm, height=5.0cm}
     \psfig{file=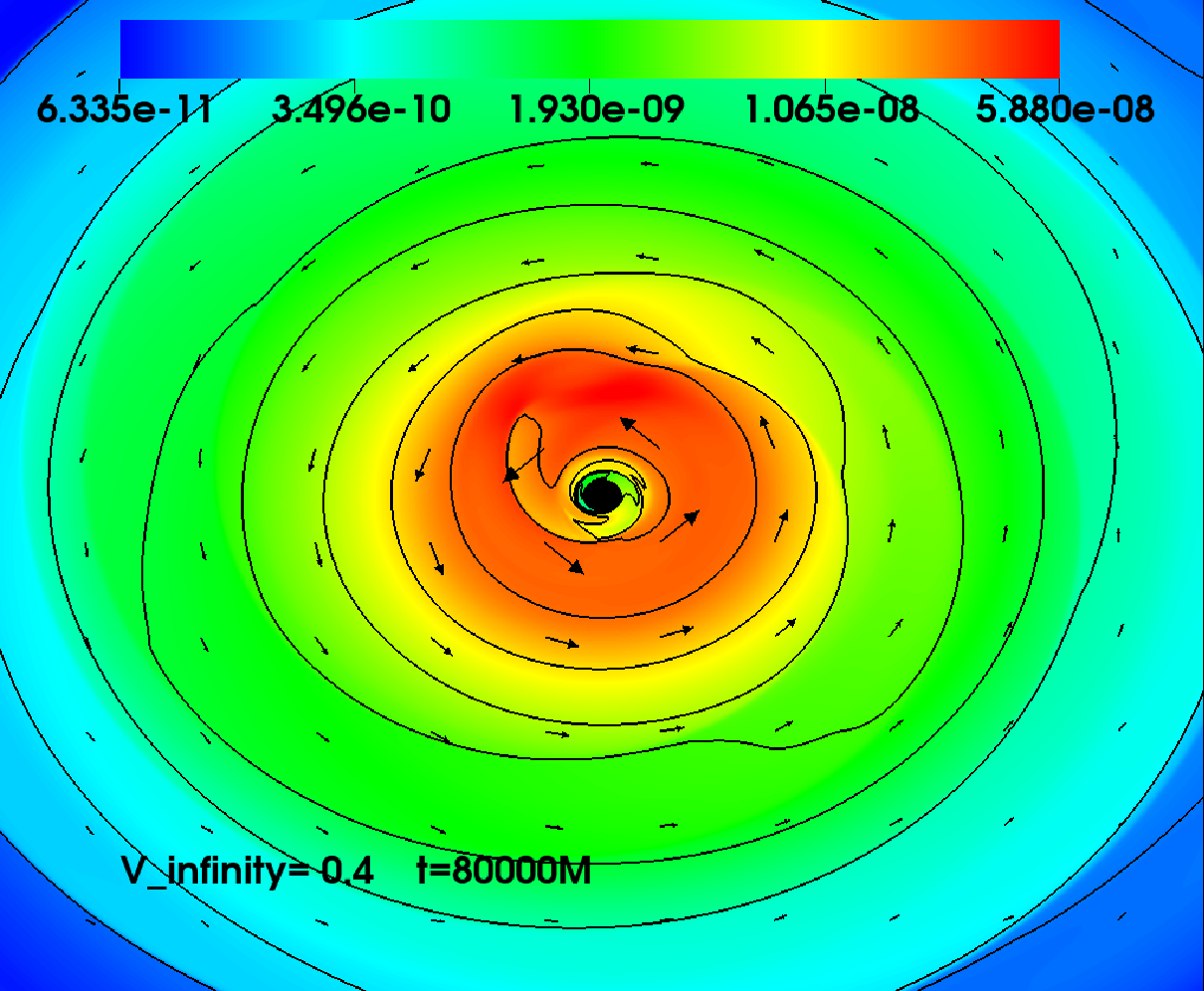,width=7.0cm, height=5.0cm} 
     \caption{In the case of a rapidly rotating black hole ($a/M=0.9$), color, contour, and vector plots of the rest mass density are given, showing the shock cone formed by BHL accretion (left column) and the final snapshot of the plasma reaching the steady state after BHL accretion is stopped (right column). Each row of screenshots corresponds to a specific value of asymptotic velocity, clearly demonstrating the complete damping of the shock cone into the black hole and the presence of a newly formed physical mechanism. The screenshots are displayed within the range of $[x_{min}, y_{min}] = [-70M, -70M]$ and $[x_{max}, y_{max}] = [70M, 70M]$.
    }
\vspace{1cm}
\label{Color_1}
\end{figure*}

Fig.\ref{Color_1} shows the variation of the rest-mass density in the equatorial plane for a spin parameter of $a/M=0.9$ at different asymptotic velocities. The graphs in the left column display the dynamic structure of the shock cone that forms around the black hole as a result of the BHL accretion, long after the system has reached the steady state. In contrast, the graphs in the right column present the dynamic structure that emerges under the same physical conditions, but long after the BHL accretion has stopped and the plasma has settled into a stable configuration.

After BHL accretion is halted at approximately $t=6000M$, the graphs in the right column of Fig.\ref{Color_1} illustrate the dynamic structure of the plasma at $t=80000M$. As can be seen from these snapshots, once the BHL accretion is stopped, the mater around the black hole either rapidly falls into the black hole or is ejected from the computational domain. As a result, compared to the case with the shock cone, the rest-mass density around the black hole significantly decreases, forming a torus-like and spiral-shaped structure. The internal dynamics of these structures vary according to the different asymptotic velocities. The differences in the newly formed plasma structure significantly affect the resulting QPO frequencies. This situation is discussed in detail in Section \ref{compare_previous}.

To analyze the outcomes of these different physical scenarios in Fig.\ref{Color_1}, we represent the rest-mass density using a color scale, with overlaid contour plots of the density and vector fields illustrating the motion of matter around the black hole. As observed in the right column, the matter generally exhibits QPOs around the black hole. These oscillations may potentially serve as an explanation for the LFQPOs and the HFQPOs observed in various astrophysical systems.

A long time after BHL accretion is completely halted, the black hole forms a torus-like plasma structure around it, leading to stable QPOs. As can also be seen in Fig.\ref{Color_1}, the spiral shock waves are observed to form in the strong gravitational field. These shock waves evolve over time after the QPOs are established, sometimes appearing as single-armed or double-armed structures.
Although the spiral arms are weak, their formation contributes to QPOs that arise in a strong gravitational field. This is because nonaxisymmetric shock (one-armed) or two-armed spiral shocks induce oscillations in the plasma \cite{2006smqw.confE.105C}. Furthermore, because of the interaction of this spiral structure with the dynamic behavior of the plasma, it excites internal waves, further contributing to the formation of QPOs \cite{2005MNRAS.362..789R, 2008MNRAS.386.2297F}. Therefore, the presence of spiral arms in a strong gravitational field causes the surrounding matter to exhibit quasi-periodic motion. Such a quasi-periodic behavior facilitates the formation of radial epicyclic frequencies.

\begin{figure*}
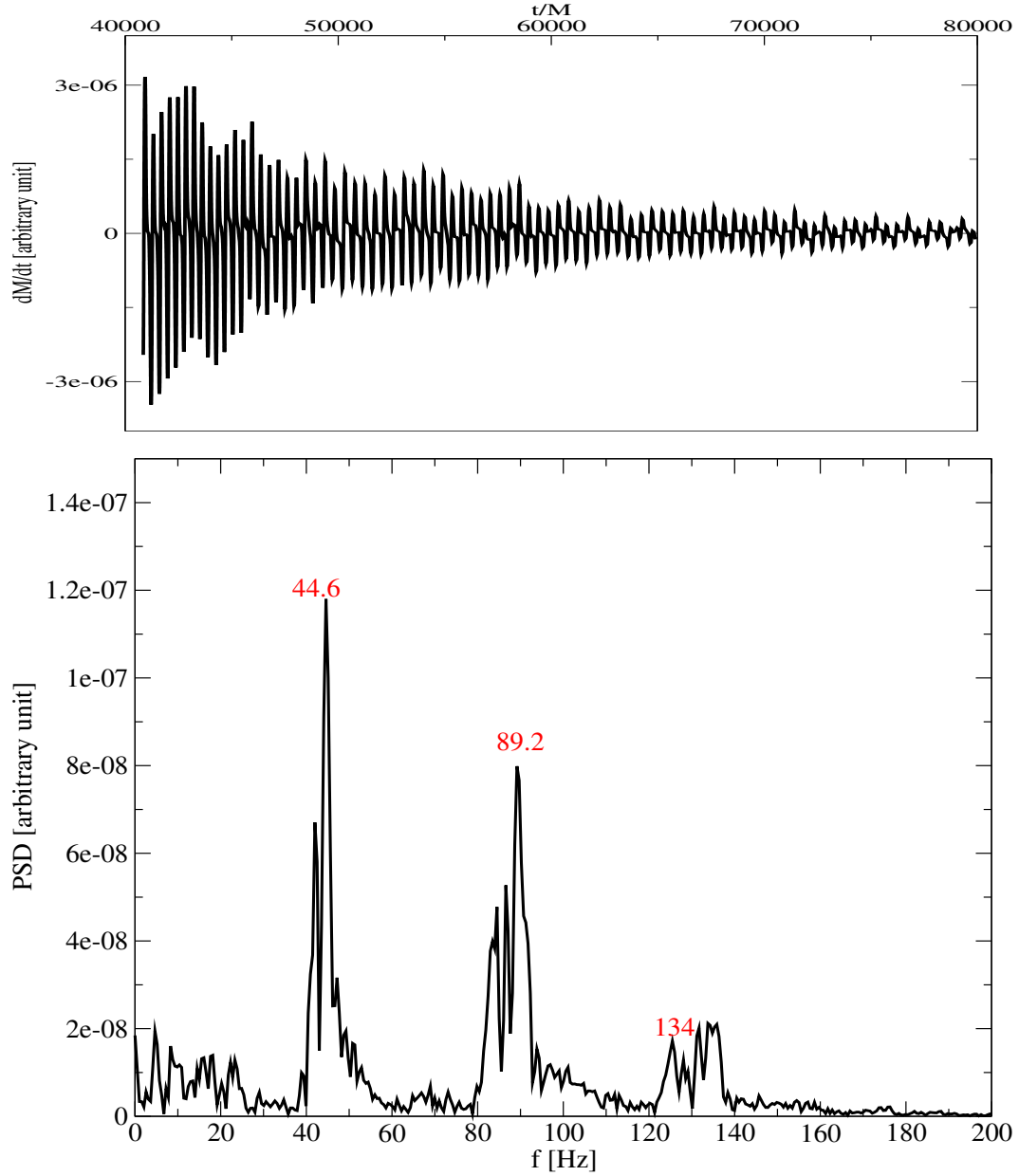

  \vspace{1cm}
  \center
  \psfig{file=MA_a09_V01_40th_80th.eps,width=14.0cm, height=6.0cm}\\
  \vspace*{0.3cm}
   \psfig{file=PSD_a09_01.eps,width=14.0cm, height=10.0cm}\\    
     \caption{Demonstration of the mass accretion rate and the PSD analysis calculated from this accretion rate in the strong gravitational region around the rapidly rotating black hole with $M=10 M_{\odot}$. For the black hole rotation parameter of $a/M=0.9$ and an asymptotic velocity of $V_{\infty}/c=0.1$, after stopping BHL accretion at $t=5000M$, the majority of matter is dumped into the black hole. The PSD analysis is performed on the mass accretion rate between $t=40000M$ and $t=80000M$, where the plasma around the black reaches a steady-state, and the resulting QPO frequencies are presented.
    }
\vspace{1cm}
\label{Mas_PSD_v01}
\end{figure*}

Studying the QPOs that arise around a black hole helps us to understand the physical mechanisms behind the LFQPOs and HFQPOs observed in different astrophysical sources. In Fig.\ref{Mas_PSD_v01}, the rest-mass density and the power spectrum analysis are performed for a black hole with $a/c=0.9$ in a scenario where the asymptotic velocity is $V_{\infty}/c=0.1$. This analysis is performed long after the BHL accretion is halted and the plasma around the black hole has reached a quasi-steady state. Long after BHL accretion has stopped, the HFQPOs of matter around the black hole are clearly observed in the mass accretion rate. As also seen in Fig.\ref{Mas_PSD_v01}, the fundamental mode frequency appears at $44.6$ Hz, with additional peaks at $89.2$ Hz and $134$ Hz. These higher frequencies result from nonlinear coupling between the fundamental mode and newly formed oscillatory modes, leading to a resonance pattern $1:2:3$. The most likely physical origin of these HFQPOs is the radial epicyclic motion.

During BHL accretion, the asymptotic velocity is set at $V_{\infty}/c=0.1$, meaning that the inflowing matter is in a sonic regime as it falls toward the black hole. Because of this, the effect of the black hole spin parameter on QPO formation is not observed. In other words, even though BHL accretion is stopped, the dominant fundamental mode remained radial epicyclic oscillations. This is because the accreting matter, in the presence of a strong gravitational field, does not experience significant interaction with the black hole which causes Lense-Thirring precession QPO frequency.

\begin{figure*}
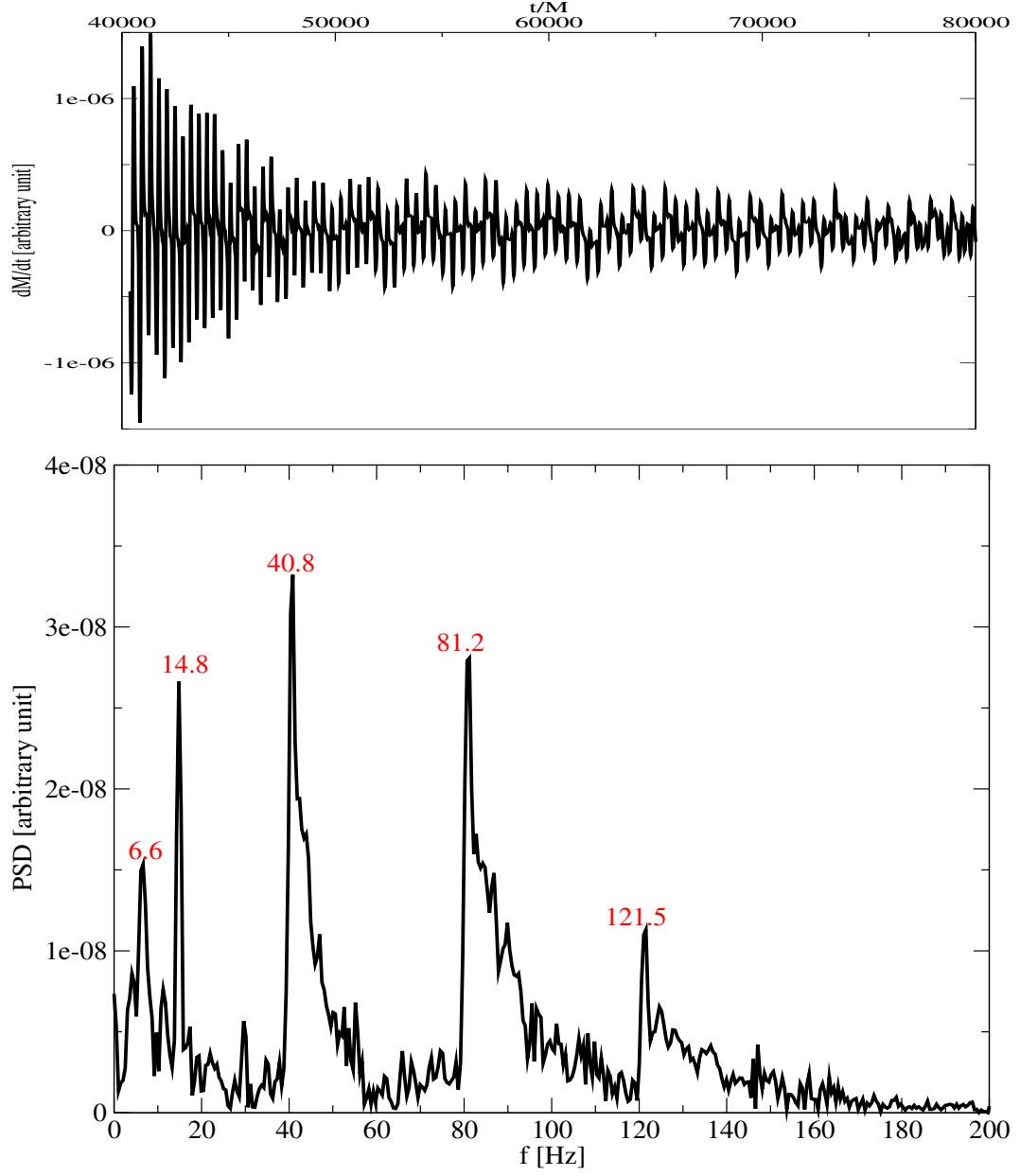

  \vspace{1cm}
  \center
  \psfig{file=MA_a09_V02_40th_80th.eps,width=14.0cm, height=6.0cm}\\
  \vspace*{0.3cm}
   \psfig{file=PSD_a09_02.eps,width=14.0cm, height=10.0cm}\\    
     \caption{Same as Fig.\ref{Mas_PSD_v01}, but it is for the QPO frequencies formed in the case of $V_{\infty}/c=0.2$.
    }
\vspace{1cm}
\label{Mas_PSD_v02}
\end{figure*}

For the same rapidly rotating black hole ($a/M=0.9$), but this time with an asymptotic velocity of $V_{\infty}/c=0.2$—meaning that the inflowing matter remains supersonic throughout the BHL accretion process—the rest-mass accretion rate analysis and power spectrum analysis are conducted long after BHL accretion is stopped. The results are presented in Fig.\ref{Mas_PSD_v02}. Similar to Fig.\ref{Mas_PSD_v01}, the fundamental HFQPOs mode appears at $40.8$ Hz, which is again attributed to radial epicyclic oscillations. Due to nonlinear coupling, additional high-frequency modes emerge at $81.2$ Hz and $121.5$ Hz, forming a resonance pattern $1:2:3$ (i.e. $40.8:81.2:121.5$).

In addition to these HFQPOs, this model also exhibits the presence of LFQPOs, appearing at $6.6$ Hz and $14.8$ Hz. These LFQPOs are believed to originate from the Lense-Thirring precession effect. When a supersonic plasma forms around the rapidly rotating black hole, its interaction with the black hole within the strong gravitational field produces LFQPOs, which are consistent with theoretical predictions. Therefore, our findings provide a possible explanation for why certain astrophysical sources, which exhibit continuous variability, can generate both LFQPOs and HFQPOs. The results presented here help to clarify the physical origin of such QPOs observed in various astrophysical environments.

While examining the behavior of matter around the black hole when BHL accretion is halted, we also modeled the shock cone formation process for the asymptotic velocities of $V_{\infty}/c=0.3$ and $V_{\infty}/c=0.4$, where the matter falls toward the black hole at a certain asymptotic speed. As the asymptotic speed increases, comparing Figs.\ref{Mas_PSD_v03} and \ref{Mas_PSD_v04} with Fig.\ref{Mas_PSD_v02}, the matter exhibits a stronger supersonic behavior. We can state that this initial supersonic behavior does not significantly affect the QPO frequencies that form after the BHL accretion is halted. When Fig.\ref{Mas_PSD_v02} is compared with Figs.\ref{Mas_PSD_v03} and \ref{Mas_PSD_v04}, the HFQPOs and their nonlinear couplings are still present. Additionally, because of the Lense-Thirring effect, LFQPOs have also emerged.

Consequently, when Figs.\ref{Mas_PSD_v01}, \ref{Mas_PSD_v02}, \ref{Mas_PSD_v03}, and \ref{Mas_PSD_v04} are compared, it is observed that for the Lense-Thirring effect to become apparent and for the QPO frequencies caused by the curvature of spacetime in the strong gravitational field of the black hole to be observed, the matter must exhibit a supersonic behavior around the black hole. The frequencies generated by the Lense-Thirring effect appear as LFQPOs. However, the radial epicyclic frequency has also been observed in both the sonic and supersonic behaviors of the plasma. These frequencies can be detected as HFQPOs by detectors. Thus, for sources $GRO J1655-40$, $GRS 1915+105$, $XTE J1550-564$, and $H1743-322$ \cite{wang20242022,BelloniMNRAS2012,Strohmayer2001ApJ,Remillard2006ARA&A,Varniere_2018,Majumder_2022,Liu_2021, Dhaka_2023}, which exhibit both the LFQPO and the HFQPO behavior, the halted BHL accretion mechanism with supersonic flow modeled here can be proposed as a physical mechanism to explain the observed QPOs.

\begin{figure*}
  \vspace{1cm}
  \center
  \psfig{file=MA_a09_V03_40th_80th.eps,width=14.0cm, height=6.0cm}\\
  \vspace*{0.3cm}
   \psfig{file=PSD_a09_03.eps,width=14.0cm, height=10.0cm}\\    
     \caption{Same as Fig.\ref{Mas_PSD_v01}, but it is for the QPO frequencies formed in the case of $V_{\infty}/c=0.3$.
    }
\vspace{1cm}
\label{Mas_PSD_v03}
\end{figure*}

\begin{figure*}
  \vspace{1cm}
  \center
  \psfig{file=MA_a09_V04_40th_80th.eps,width=14.0cm, height=6.0cm}\\
  \vspace*{0.3cm}
   \psfig{file=PSD_a09_04.eps,width=14.0cm, height=10.0cm}\\    
     \caption{Same as Fig.\ref{Mas_PSD_v01}, but it is for the QPO frequencies formed in the case of $V_{\infty}/c=0.4$.
    }
\vspace{1cm}
\label{Mas_PSD_v04}
\end{figure*}

\subsection{Analysis of Black Holes with $a/M=0.5$ and $a/M=0.0$ }
\label{Other_Num}

\begin{figure*}
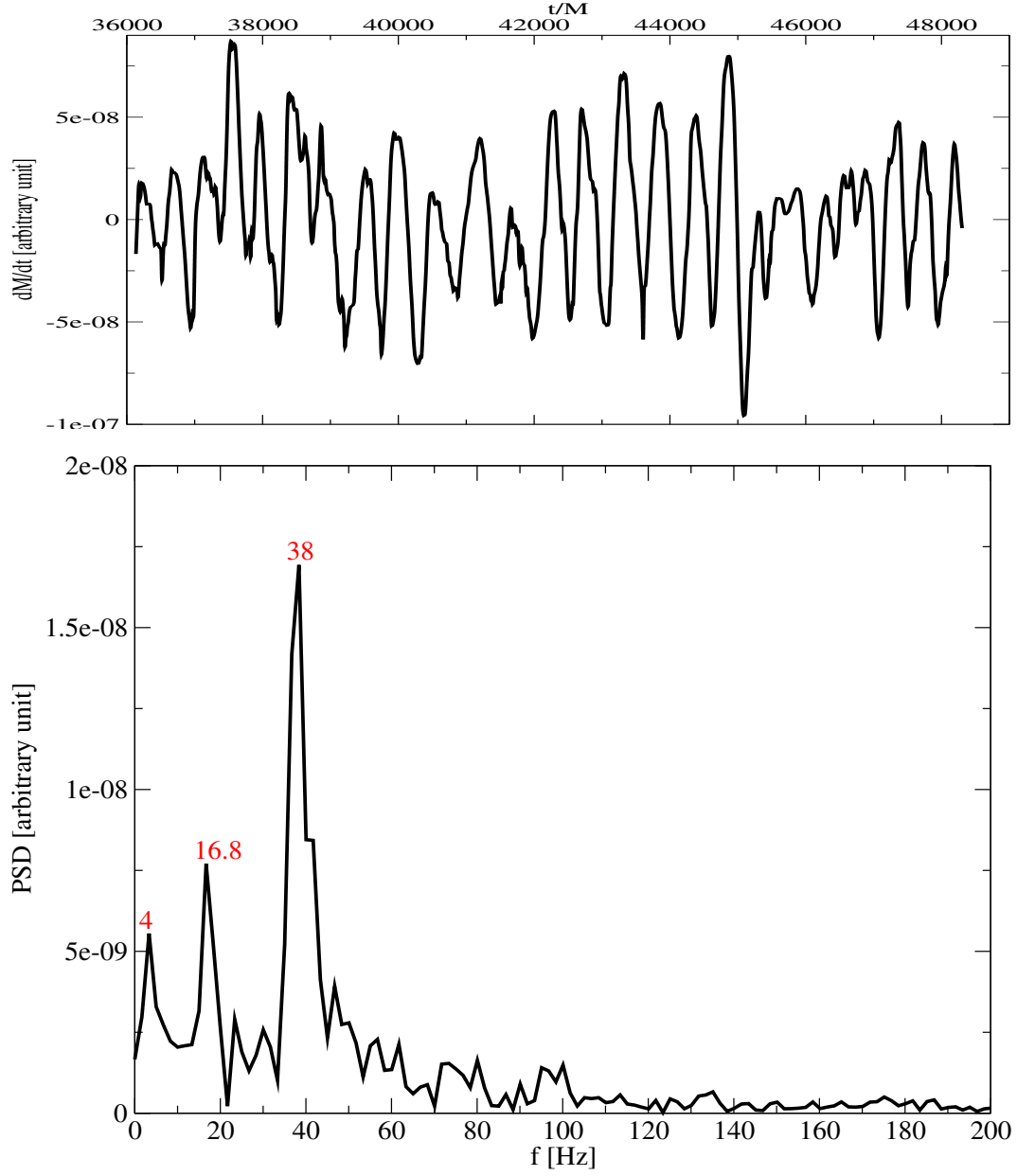

  \vspace{1cm}
  \center
  \psfig{file=MA_a05_V02_36th_48th.eps,width=14.0cm, height=6.0cm}\\
  \vspace*{0.3cm}
   \psfig{file=PSD_a05_02_36th_48th.eps,width=14.0cm, height=10.0cm}\\    
     \caption{As shown in Fig.\ref{Mas_PSD_v01}, the mass accretion rate and PSD analyses have been plotted for $a/M=0.9$ and $V_{\infty}/c=0.2$. However, this time, in the case of a slowly rotating black hole, i.e., $a/M=0.5$, PSD analysis is conducted when the matter immediately exhibited the steady-state, and the resulting QPO frequencies are presented.
    }
\vspace{1cm}
\label{Mas_PSD_a05_v02}
\end{figure*}

Since the black hole spin parameter affects the curvature of spacetime, it also influences the Lense-Thirring frequency. In this section, the effect of the spin parameter on both LFQPOs and HFQPOs has been analyzed. To examine this effect, Fig.\ref{Mas_PSD_a05_v02} presents the mass accretion rates occurring at the spin parameter $a/M=0.5$ under the same asymptotic velocity, $V_{\infty}/c=0.2$ and the power spectrum density analysis generated from this accretion rate. In Fig.\ref{Mas_PSD_a05_v02}, similar to Fig.\ref{Mas_PSD_v02}, an asymptotic velocity of $V_{\infty}/c=0.2$ is used, but in this case, a moderately spinning black hole with $a/M=0.5$ is considered. As in the case of the moderately spinning black hole in Fig.\ref{Mas_PSD_a05_v02}, LFQPOs arising due to the Lense-Thirring effect occur at $4$ Hz and $16.8$ Hz, while HFQPOs at $38$ Hz are generated due to the influence of radial epicyclic motion. The QPO frequencies observed in Fig.\ref{Mas_PSD_a05_v02}, within a certain error margin, exhibit similarities to the peaks in Fig.\ref{Mas_PSD_v02}.

However, unlike Fig.\ref{Mas_PSD_v02}, the high-frequency peak in Fig.\ref{Mas_PSD_a05_v02}, does not generate additional high-frequency peaks through nonlinear couplings. Consequently, we can infer that the spin parameter of the black hole plays a role in the emergence of nonlinear couplings. This result is consistent with studies in the literature \cite{2014CQGra..31b5023M, 2020PhRvL.125w1101D, 2023PhRvD.108l4048E}. There are several reasons for the emergence of the nonlinear coupling effect. These include the suppression or amplification of nonlinear oscillation modes during black hole-matter interactions, the formation of a new spin-orbit coupling in a strong gravitational field due to chaotic behavior \cite{2014CQGra..31b5023M}, and the occurrence of the scalarization process due to the resulting tachyonic instability \cite{2020PhRvL.125w1101D}. Additionally, the spin of the black hole influences the angular momentum and energy flux of matter, leading to the nonlinear coupling effect \cite{2013arXiv1310.1602W}.

\begin{figure*}
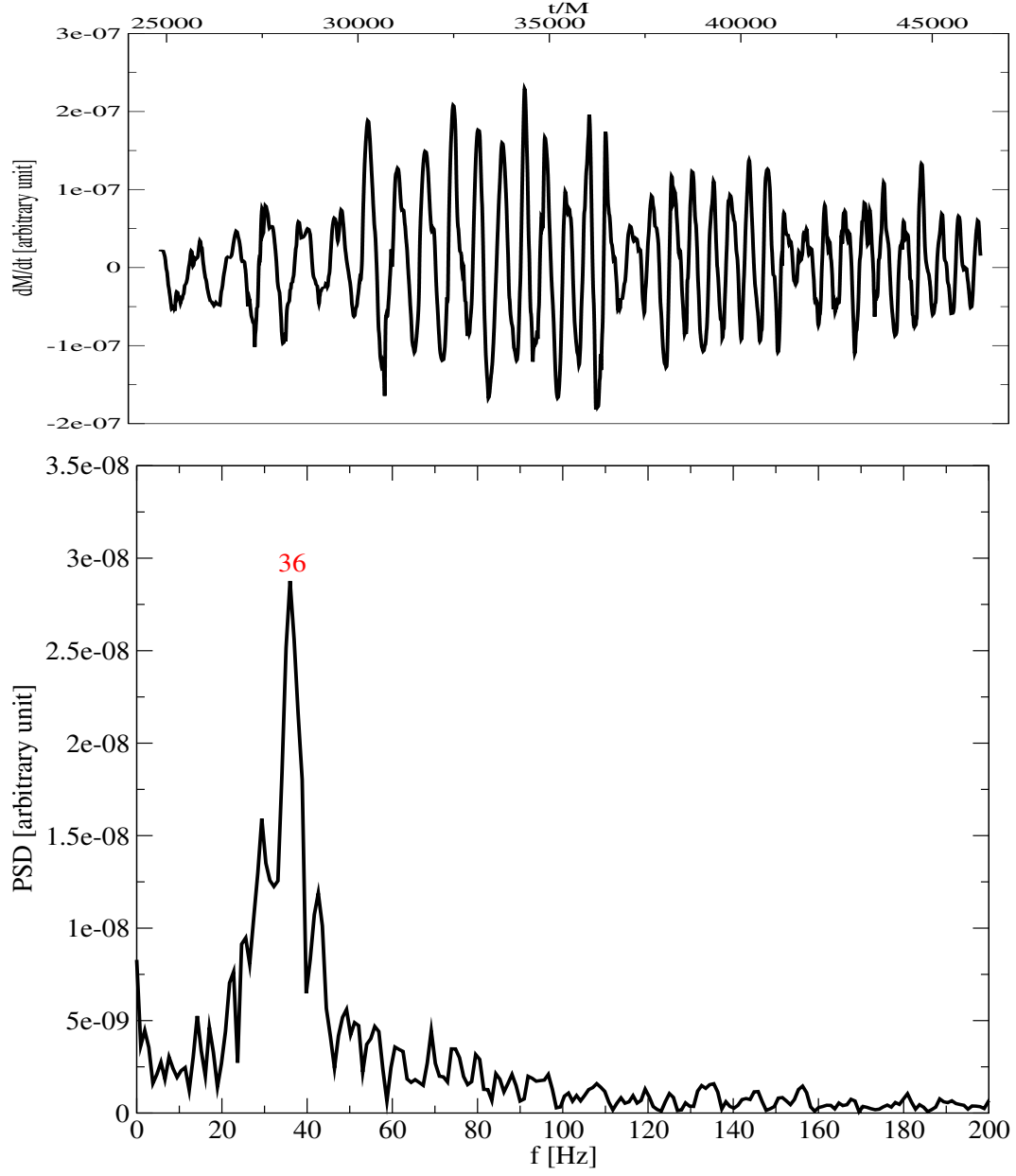

  \vspace{1cm}
  \center
  \psfig{file=MA_a0_V02_25th_46th.eps,width=14.0cm, height=6.0cm}\\
  \vspace*{0.3cm}
   \psfig{file=PSD_a0_02_25th_46th.eps,width=14.0cm, height=10.0cm}\\    
     \caption{Same as Fig.\ref{Mas_PSD_a05_v02}, but it is for the  non-rotatng black hole, $a/M=0.0$.
    }
\vspace{1cm}
\label{Mas_PSD_a00_v02}
\end{figure*}

To reveal the effect of the black hole spin parameter on the generated QPOs, we analyze the QPOs formed around the non-rotating black hole. For this purpose, similar to Figs.\ref{Mas_PSD_v02} and \ref{Mas_PSD_a05_v02}, the mass accretion rate and the power spectrum analysis for the non-rotating black hole at the same asymptotic velocity of $V_{\infty}/c=0.2$ are presented in Fig.\ref{Mas_PSD_a00_v02}. As seen in Fig.\ref{Mas_PSD_a00_v02}, no LFQPOs are observed, while only the HFQPO at $36$ Hz is detected. Additionally, since the black hole does not rotate, the high-frequency peaks arising from nonlinear coupling, which are observed in Figs.\ref{Mas_PSD_v02}, \ref{Mas_PSD_v03}, and \ref{Mas_PSD_v04}, do not form. The reason for the absence of nonlinear couplings is the lack of a physical mechanism that could excite the fundamental HFQPO mode and generate new frequencies. These physical mechanisms could include shock cones forming around the black hole, strong spiral shock waves, or the warped spacetime caused by the black hole rotation \cite{2008MNRAS.386.2297F}.

As a result, it has been demonstrated that the black hole spin parameter significantly affects the QPO frequencies. If the plasma exhibits supersonic behavior around the rapidly rotating black hole, both LFQPOs and HFQPOs form. However, as the spin parameter decreases, LFQPOs continue to emerge, while the frequencies caused by nonlinear coupling disappear. In the case of a nonrotating black hole model, as expected, both LFQPOs and the frequencies originating from nonlinear coupling completely vanish if there is no physical mechanism to excite the QPO modes.

\section{Effect of Spin and Asymptotic Velocity on LFQPOs and HFQPOs}
\label{low_high_QPOs}

The identification of the physical origins of LFQPOs and LFQPOs around the black holes and their consistency with observations contribute to understanding the physical properties of the black holes. This is because QPOs emerging in strong gravitational fields provide insight into black hole mass, spin parameter, accretion flow dynamics, and relativistic effects. Theoretically, it is well known that the Lense-Thirring effect produces LFQPOs \cite{Stella1998ApJ, Ingram2009MNRAS, Wu2023PhRvD}. This occurs because the spin parameter of the black hole induces a precessional motion in the surrounding matter near the event horizon. Consequently, higher-spin black holes create a stronger frame-dragging effect, which results in shifting the QPO frequencies upward. As shown in Table \ref{QPO_table}, this behavior has also been confirmed numerically. When $a/M=0.9$, LFQPOs appear, whereas for $a/M=0.0$, these frequencies disappear. In addition to this, the asymptotic velocity that drives matter toward the black hole in the BHL accretion also affects QPO formation. LFQPOs are observed at higher asymptotic velocities. In other words, if the matter around the black hole exhibits supersonic motion, LFQPOs are generated.

HFQPOs, on the other hand, originate from the orbital motion of matter. More specifically, the radial epicyclic motion of matter contributes to these frequencies. Since the motion of matter is influenced by the black hole spin, a faster-spinning black hole results in higher HFQPO values. As confirmed numerically in Table \ref{QPO_table}, as the black hole spin increases from $a/M=0.0$ to $a/M=0.9$, the QPO frequency increases from $36$ Hz to $40.84$ Hz. Furthermore, as seen in Table \ref{QPO_table}, HFQPOs can also generate new frequencies through nonlinear couplings. The presence of nonlinear couplings indicates complex interactions between different modes of disk oscillations. Table \ref{QPO_table} shows that for high-spin cases ( $a/M=0.9$), nonlinear couplings appear, but they vanish for moderate or zero spin. This suggests that nonlinear interactions in the inner accretion disk are strongly influenced by the black hole spin and the accretion geometry and it is also consistent with the existence literature \cite{2013PhRvL.111d1101S, 2024PhRvD.109j1503R}.

There are no studies in the literature that directly demonstrate a direct relationship between accretion velocity and QPOs. However, previous research has shown that since the accretion velocity significantly influences the formation of resonance conditions and the accumulation of matter toward the black hole during black hole-matter interactions, it also affects QPO formation and its frequency \cite{Mukhopadhyay_2009, 2009MNRAS.398.1105R}. Therefore, the results obtained here are consistent with the literature, and it is shown in our numerical simulations that the accretion velocity plays a role in QPO formation, particularly in determining LFQPOs and whether HFQPOs emerge due to nonlinear couplings.

As seen in Table \ref{QPO_table}, for $a/M=0.0$, neither LFQPOs nor nonlinear couplings are observed. LFQPOs are commonly attributed to Lense-Thirring precession, which occurs due to frame-dragging around the spinning black hole \cite{2020NatPh..16.1069C, 2020PhRvD.102b4021K}. Since there is no frame-dragging at $a/M=0.0$, these frequencies do not form.  Additionally, LFQPOs can also arise from the precession of the inner accretion disk. However, for $a/M=0.0$, no torque is transferred from the black hole to the surrounding matter, preventing the formation of this precession. At the same time, without spin, the inner disk does not experience strong differential frame-dragging, reducing nonlinear interactions.

\begin{table}
\footnotesize
\caption{
The QPO frequencies obtained from numerical simulations, their occurrence based on the black hole spin parameter and asymptotic velocity, and the possible nonlinear coupling probabilities are presented.
}
 \label{QPO_table}
\begin{center}
  \begin{tabular}{ccccc}
    \hline
    \hline

    $a/M$ & $V_{\infty}/c$   & LFQPOs (Hz) & HFQPOs (Hz) & Nonlinear Couplings (Hz)  \\
          &                &            & (Fundamental Mode)  &   \\   
    \hline
    $0.9$ & $0.1$& $No$          & $44.6$ & $89.2$, $134$      \\
    $0.9$ & $0.2$& $6.6$, $14.8$ & $40.8$ & $81.2$, $121.5$     \\
    $0.9$ & $0.3$& $1.5$, $22.62$& $49$   & $97.3$, $145.6$     \\
    $0.9$ & $0.4$& $7.4$         & $43$   & $86$,   $128.5$     \\
    \hline
    $0.5$ & $0.2$& $4$, $16.8$   & $38$   & $No$                 \\
    $0.0$ & $0.2$& $No$          & $36$   & $No$                  \\
 
    \hline
    \hline
  \end{tabular}
\end{center}
\end{table}

As shown in Table \ref{QPO_table}, the QPO frequencies correspond to global modes that are independent of the chosen time window of the data, can be identified throughout the disk, and are not sensitive to the specific diagnostic parameter used. This confirms that the QPOs observed in our simulations are not numerical artifacts but are instead the natural outcome of excited oscillation modes in the disk. The first test we performed is to divide the accretion rate signal,  after the BHL accretion is halted and the disk reached a quasi-steady state, into multiple windows, and recompute the PSD in each segment. In all cases, almost the same frequencies reappeared. A more detailed version of this test was presented in our earlier work \citep{2024EPJC...84..524D, Donmez2023arXiv231013847D, Donmez2024Submitted}. In all of our numerical analyses, the power spectrum peaks are found at nearly identical frequencies. The second test, frequently applied in our analysis, is to compute the PSD at different radial locations in the flow (e.g., $r=2.3M$, $r=6.1M$, and $r=12M$). As seen in previous references \citep{Donmez6, Donmez2024Univ, Donmez2024arXiv240701478D, Donmez2024arXiv240216707D} and confirmed again in this study, the peaks appear at nearly the same frequencies regardless of radius, with only their amplitudes varying depending on the plasma density near the black hole. This indicates that the peaks originate from global disk instabilities rather than local turbulence, demonstrating that the QPOs are global rather than local in nature \citep{Donmez6, 2011MNRAS.417.2899Z, Donmez5}. In addition, beyond calculating the PSD from the mass accretion rate as we did throughout this paper, we also performed the analysis using the mass density at a fixed location near the black hole, sampled at different times. We found that the resulting frequencies are essentially identical to those obtained from the accretion rate. This further confirms that the frequencies are not artifacts of the numerical method. Together, these tests demonstrate that the frequencies reported in Table \ref{QPO_table} are robust against variations in temporal sampling, radial location, and diagnostic quantity. This strengthens the reliability of our numerical results and places them on firmer ground for comparison with observed QPOs in X-ray binaries \citep{Remillard2006ARA&A, Ingram2009MNRAS}.


\section{Comparison of Numerical QPO Frequencies with Observational Data}
\label{comparision}

We compare the QPO frequencies obtained from the analysis of the precession motion of plasma in the strong gravitational field after the BHL accretion is halted, as given in Table \ref{QPO_table}, with the observational results presented in Table \ref{Observational_QPO_table}. As noted in the final paragraph of the Introduction, in this paper we adopt geometrized units. In geometrized units, the frequency has units of $1/M$, where $M$ is the black hole mass. Since the frequencies we compute from the PSD analyses depend explicitly on the black hole mass, the results obtained here can be used not only to explain the QPO frequencies observed around stellar-mass black holes but also to shed light on the observational findings for supermassive black hole sources. Therefore, in order to compare the numerical QPO frequencies given in Table \ref{QPO_table} with the observational results summarized in Table \ref{Observational_QPO_table}, we must convert our numerical results from geometrized units ($1/M$) to SI units (Hz). The conversion from geometrized to SI units is performed using $1/M = c^3/(GM)$ where $c$ is the speed of light and $G$ is the gravitational constant. As shown in Table \ref{Observational_QPO_table}, since our comparisons in this study are mainly with QPO frequencies observed around stellar-mass black holes and the typical masses of these sources are around $M=10M_{\odot}$, we adopt $M=10M_{\odot}$ in carrying out this conversion. In this case, the conversion formula from geometrized to SI units becomes,

    \begin{eqnarray}
    \nu_{new}= \nu_{old}\; \frac{10M_{\odot}}{M_{BH}}.
    \label{Scaling}
  \end{eqnarray}    

Using Eq.\ref{Scaling}, we can also recalculate our numerical frequencies for massive black holes and make a direct comparison with any given sources. For the rapidly spinning black hole case ($a/M=0.9$), our low-frequency numerical results show agreement with the observed $Type-C$ QPOs of $GRS 1915+105$ and $XTE J1550-564$. However, the numerically obtained frequencies of $36$ Hz, $38$ Hz, $40.8$ Hz, and $44.6$ Hz show consistency with the observed HFQPO frequencies of $GRS 1915+105$. Furthermore, we can compare our findings with the AGN source $RE J1034+396$, which is known to host a massive black hole. In this case, if we rescale the frequencies found in the PSD analyses for the high-mass source $REJ1034+396$ (with a mass of $10^6M_{\odot}$) given in Table 3, we obtain $\nu_{SMBH}= \frac{36}{10^5} = 3.6x10^{-4}$ Hz and $\nu_{SMBH}= \frac{40.8}{10^5} = 4.08x10^{-4} Hz$. The rescaled frequencies align well with the observed frequency given in Table \ref{Observational_QPO_table} for $RE J1034+396$.

  Overall, the numerical results indicate that LFQPOs only emerge in the case of supersonic accretion when the black hole has a high spin parameter. This finding is in agreement with the numerical simulations and observations \cite{Ingram2009MNRAS, 2015MNRAS.452.3451C, 2020MNRAS.492..804M, 2022MNRAS.510..807S}. In other words, strong LFQPOs are found to form in the plasma surrounding rapidly rotating black holes, which is also consistent with observational data. For HFQPOs, the resonance structure ($3:2$, $2:1$) observed in the numerical simulations due to nonlinear couplings perfectly matches the observational results \cite{2015A&A...581A..35B, 2022JCAP...08..034B}. However, some of the HFQPO frequencies observed in certain sources (e.g. $300$ Hz, $450$ Hz in $GRO J1655-40$) do not appear in the numerical simulations. One reason for this may be the assumptions and simplifications made in setting up the simulations. In this study, we neglect the effects of magnetic fields and numerically solve the $2D$ GRHD equations to model the plasma structure formed by matter acting as an ideal gas around the black hole, along with the resulting shock waves. Under these assumptions, LFQPOs and some HFQPOs can indeed be recovered numerically, but certain frequencies reported in observations do not appear in the simulations. In contrast, the higher-frequency QPOs (e.g., $300–450$ Hz) may require additional physical mechanisms not captured here, such as magneto-rotational turbulence, disk inhomogeneities, or strong magnetic stresses. These processes can enhance the coupling between the inner disk and the black hole spin. Furthermore, our analysis assumes a black hole mass of $M=10M_{\odot}$, and while scaling relations allow comparisons across systems, uncertainties in the true mass and spin of $GRO J1655-40$ introduce further shifts in the predicted frequencies. Furthermore, because the accretion mechanism in this study is modeled in the equatorial plane, the vertical epicyclic frequencies are neglected, which may also explain why the $300$ and $450$ Hz features do not emerge in our simulations. Therefore, the absence of the $300$ Hz and $450$ Hz frequencies in our results does not imply that the mechanism modeled here is incapable of producing them, but rather that the restricted hydrodynamic setup isolates only a subset of the physically relevant oscillatory modes. A complete $3D$ GRMHD treatment of the problem considered in this paper would allow us to investigate whether the mechanisms mentioned above contribute to the formation of these HFQPOs. Lastly, in recent years, alternative theories of gravity have been proposed that may produce results capable of explaining the origin of these frequencies. If such a scenario holds, these HFQPOs could serve as important observational probes for testing strong gravity.

\begin{table}
\footnotesize
\caption{
QPO frequencies of different sources found based on the X-ray data obtained from observational results.
}
 \label{Observational_QPO_table}
\begin{center}
  \begin{tabular}{ccccc}
    \hline
    \hline

    Source & Mass $(M_{\odot})$   & LFQPOs (Hz) & HFQPOs (Hz) & Reference  \\
    \hline
    $GRS 1915+105$  & $12–18$ & $1–10$     & $41, 67$      &  \cite{Strohmayer2001ApJ, Misra2004, Liu_2021, Majumder_2022, Motta2023}    \\
    $GRO J1655-40$  & $6.3$   & $0.1–10$   & $300, 450$    &  \cite{Remillard1999ApJ, 2012MNRAS.427..595M, Motta2014MNRAS}  \\
    $XTE J1550-564$ & $9.1$   & $0.1–10$   & $184, 276$    &   \cite{Varniere_2018}\\
    $H1743-322$     & $8–10$  & $0.1–10$   & $240$         &   \cite{2006ApJ...637.1002R, 2012ApJ...754L..23A} \\
    $RE J1034+396$  & $10^6$  &  $2.7 \times 10^{-4}$ &  &  \cite{Czerny2016AA, Jinmnras2020}  \\
    \hline
    \hline
  \end{tabular}
\end{center}
\end{table}
%


\section{New Insights into QPO Formation: A Comparison with Earlier Findings}
\label{compare_previous}

Torus-like structures form around the black holes as a result of various astrophysical scenarios. Studying these structures is crucial for understanding accretion physics, relativistic effects, and the QPO behavior induced by the torus. Previously, by numerically solving the general relativistic hydrodynamic equations, we investigated the dynamic structure and physical properties of the tori around the Schwarzschild and the Kerr black holes \cite{Donmez4, 2017MPLA...3250108D, Donmez_Turkish_Journal, 2019PAIJ....3..184D}. Since a torus is a stable structure, it does not inherently generate QPO frequencies. However, perturbation of the torus around a black hole is a natural outcome of astrophysical processes. For instance, matter captured by the strong gravitational field of the black hole can fall toward it, perturbing the surrounding torus. This scenario was studied in \cite{Donmez4}, where numerical analyzes demonstrated that a perturbed torus undergoes instabilities, leading to the formation of HFQPOs (as presented in Table 2 of \cite{Donmez4}). Although \cite{Donmez4} showed only the formation of HFQPOs, in this study, as seen in Table 2, we find the emergence of both LFQPOs and HFQPOs. Therefore, the results of this study are in direct agreement with the observed QPO frequencies in microquasars.

The results we obtained here are important for comparing the QPO frequencies generated by the oscillation of the shock cone around the Kerr black hole and the modes trapped inside the cone with BHL accretion. Previously, in \cite{Donmez6}, the dynamic structure of the shock cone formed around the Kerr black hole was examined under both sonic and supersonic flow conditions, depending on different black hole spin parameters. The QPO frequencies obtained are provided in Table 2 of \cite{Donmez6}. As seen in this table, only HFQPOs are formed. In contrast, in this study, by halting BHL accretion, a new plasma structure around the black hole has formed, resulting in both LFQPOs and HFQPOs. When the results of these two studies are compared with the sources given in Section \ref{comparision}, the QPO frequencies obtained in this study are consistent because they explain both the LFQPO and HFQPO frequencies. Therefore, the plasma structure that forms in the strong gravitational field around the black hole because of the cessation of the BHL mechanism sheds light on the physical mechanisms that can be proposed for QPOs in the literature. On the other hand, the fundamental HFQPO modes and their nonlinear coupling-generated frequencies in Table 2 of \cite{Donmez6}, as well as the increase in these frequencies with the black hole spin, are in perfect agreement with the results obtained in this study. In conclusion, for QPO formation in observed sources, the best physical model that can be proposed is the plasma structure found in this study. The behavior of this plasma, influenced by both precession motion and strong rotation, provides a direct match with observations, ensuring the formation of QPOs.

The decrease in the rest-mass density of matter (plasma) around the black hole and the change in its dynamic structure can affect the trapping or excitation of waves that may occur within the plasma. At the same time, it can alter the nonlinear coupling mechanism, leading to the formation of weaker and fewer QPOs in this scenario \cite{2008MNRAS.386.2297F}. However, because of these plasma changes, the variation in the density of the Comptonizing cloud alters the cooling and infall timescales, which in turn can contribute to the formation of LFQPOs \cite{2020MNRAS.492..804M}. When comparing the numerical PSD analyzes obtained in Section \ref{NumRes} with those of \cite{Donmez6, Donmez4}, it is observed that the resulting QPO frequencies are smaller in number and have lower amplitudes. Furthermore, these plasma variations have caused changes in the resonant states, frequencies, and their amplitudes \cite{2004PASJ...56..931K}. As a result, the numerical findings obtained are consistent with those in the literature and provide an explanation for the HFQPO and LFQPO observed from different sources.


\section{Conclusions}
\label{Conclusions}

In this paper, by uncovering the impact of the newly emerged physical mechanism, formed as a result of halting BHL accretion around a Kerr black hole, on the generation of QPOs, we link plasma dynamics to both LFQPOs and HFQPOs within a unified framework. To achieve this goal, we model the complete disappearance of the shock cone formed around the black hole. The shock cone is a physical mechanism previously established in strong gravitational fields around the black holes using the BHL mechanism in both general relativity and alternative gravity theories. We examine how this shock cone mechanism evolves after the cessation of BHL accretion and observe the formation process of the newly emerging plasma structure. Eventually, after a long dynamical evolution, the matter in plasma reaches a steady-state configuration around the black hole, forming a quasi-periodically oscillating plasma structure. We investigate how the dynamic evolution of the plasma structure and the resulting precession motion depend on the black hole spin parameter and asymptotic velocity. Finally, by performing a power spectrum analysis, we identify the QPO frequencies that emerge from the system.

The structure of the plasma formed by completely halting BHL accretion at different asymptotic velocities, where matter falls toward the black hole, has been revealed. Contour plots clearly show that the plasma exhibits QPOs around the black hole. This quasi-periodic behavior facilitates the precessional motion of matter around the black hole. Power spectrum analyses indicate that QPO frequencies emerge due to the interaction between the black hole and plasma in the strong gravitational region, as well as radial precessional motion.

It has been observed that the asymptotic velocity used in modeling the shock cone around a black hole via BHL accretion affects the plasma structure formed after BHL is halted, and consequently, the resulting QPOs. For a rapidly rotating black hole with $a/M=0.9$ and an asymptotic velocity of $V_{\infty}/c=0.1$, meaning that the flow is sonic, LFQPOs are not observed. However, LFQPOs appeared at other asymptotic velocities. Therefore, for LFQPOs to be observable, the Lense-Thirring effect, which arises solely from the black hole’s rapid rotation and its warping of spacetime, would not be sufficient. By contrast, the literature notes that different mechanisms around black holes, for instance, disk truncation \citep{Ingram2009MNRAS}, have demonstrated that LFQPOs can also arise in subsonic flows. Thus, our results should be regarded as complementary rather than exclusive. Additionally, it is known that indirectly the velocity of the gas and, consequently, its angular momentum contribute to the formation and regulation of QPOs \cite{Mukhopadhyay_2009, 2009MNRAS.398.1105R}. However, based on the numerical simulations, it has been found for the first time that if the plasma exhibits supersonic behavior, LFQPOs emerge.

In a strong gravitational field, the spin parameter of the black hole significantly influences the physical phenomena that occur. As observed in this study, LFQPOs have formed in the supersonic plasma surrounding the rotating black hole, in agreement with theoretical predictions. When the black hole is non-rotating, LFQPOs are not observed, which is also fully consistent with theory. On the other hand, the fundamental HFQPO modes have emerged for both rotating and non-rotating black hole models. This demonstrates that regardless of whether the plasma exhibits sonic or supersonic motion and whether the black hole is spinning or not, the fundamental HFQPO mode is always present in a strong gravitational field. Furthermore, in agreement with theoretical expectations, the frequency of this fundamental mode increases as the black hole spins faster.

As seen in numerical simulations and Table \ref{QPO_table}, in addition to the fundamental HFQPOs, other HFQPOs have been observed that generate different resonance modes (e.g. $3:2$ and $2:1$), which are consistent with observations. These frequencies result from the fundamental mode itself and the nonlinear couplings that form between subsequently generated QPOs. Furthermore, as shown in Table \ref{QPO_table}, these couplings occur only when the black hole is rapidly rotating. Thus, the presence of nonlinear couplings in rapidly rotating black holes suggests that additional oscillatory modes are introduced due to the strong frame-dragging effect and the differential rotation of the accretion disk.

The results obtained here show consistency with the observations. In microquasars $GRS 1915+105$, $XTE J1550-564$, and $GRO J1655-40$, the observed QPO frequencies range between $30-200$ Hz. The numerical simulations presented in this study yield HFQPO frequencies in the range of $36-49$ Hz, which are consistent with black holes of approximately stellar mass $M\sim10M_{\odot}$. Additionally, the fact that the numerical simulations show double and triple the fundamental frequency suggests that the $3:2$ and $2:1$ resonance states observed in black hole QPOs can be explained through these results. This supports the idea that QPOs in black hole systems may arise from parametric resonances in the accretion disk. On the other hand, the observed frequency range for LFQPOs in black hole $X-$ray binaries is $0.1–30$ Hz, whereas the numerical simulations in this study produced LFQPO frequencies in the range of $1.5-22.6$ Hz, which aligns well with observations. However, LFQPOs did not appear when $V_{\infty}/c=0.1$ for $a/M=0.9$. This indicates that LFQPOs depend on the accretion state, reinforcing their consistency with observational data.

Furthermore, the frequencies obtained from the numerical simulations also show agreement with the theoretical models. The simulated QPO frequencies and their scaling with black hole spin are in line with the relativistic recession model (RPM) \cite{2022MNRAS.517.1469M}. In RPMs, HFQPOs correspond to orbital and periastron precession frequencies. The absence of LFQPOs for $V_{\infty}/c=0.1$ may suggest that disk truncation suppresses the precession needed for their formation \cite{2024MNRAS.528.1142B}. Moreover, the finding that LFQPO formation requires the plasma to exhibit supersonic behavior confirms that shock wave instabilities play an active role in triggering QPOs.

We will continue working in the future to uncover the effects of the black hole spin parameter, accretion dynamics, and the physical parameters governing these processes, as well as the behavior of plasma in a strong gravitational field. In this study, magnetic fields and vertical instabilities are neglected; however, in future work, we plan to incorporate modified gravity theories, magnetic fields, and full 3D simulations to examine how these additional physical effects influence the QPO frequencies identified here. We will utilize modified gravity theories to investigate the influence of these alternative gravitational models on the QPO frequencies that we have identified in this study. Through this approach, we aim to determine the impact of the scalar field, which represents the existence of dark matter, on the results obtained here.


\section*{Acknowledgments}
All simulations were performed using the Phoenix High
Performance Computing facility at the American University of the Middle East. The author respectfully extends his thanks to the referee for their critiques, which have enhanced the quality and clarity of the manuscript.
(AUM), Kuwait.\\

\newpage

\bibliographystyle{JHEP} 
\bibliography{paper.bib}

\end{document}